\documentclass[titlepage,11pt]{article}

\pdfoutput=1

\usepackage{amssymb,amsmath,amsfonts}
\usepackage{bm}
\usepackage{epsfig}
\usepackage{ccaption}
\usepackage{graphicx}
\usepackage{float}
\usepackage{breqn}

\usepackage[
      colorlinks=true,
      linkcolor=blue,
      urlcolor=blue,
      filecolor=black,
      citecolor=red,
      pdfstartview=FitV,
      pdftitle={},
        pdfauthor={Roberto Emparan},
        pdfsubject={},
        pdfkeywords={},
        pdfpagemode=None,
        bookmarksopen=true
      ]{hyperref}

\newcommand{\beq}{\begin{equation}}
\newcommand{\eeq}{\end{equation}}
\newcommand{\beqa}{\begin{eqnarray}}
\newcommand{\eeqa}{\end{eqnarray}}
\newcommand{\bea}{\begin{eqnarray}}
\newcommand{\eea}{\end{eqnarray}}

\newcommand{\pd}{\partial}
\newcommand{\nn}{\nonumber}

\newcommand{\ie}{{\it i.e.,\,}}
\newcommand{\eg}{{\it e.g.,\,}}

\newcommand{\lp}{\left(}
\newcommand{\rp}{\right)}

\newcommand{\ord}[1]{{\mathcal O}\lp #1\rp}

\newcommand{\sR}{\mathsf{R}}

\newcommand{\hr}{\hat{r}}
\newcommand{\htt}{\hat{t}}

\newcommand{\hph}{\hat{\phi}}
\newcommand{\br}{\rho}

\newcommand{\nh}{m_\phi}
\newcommand{\kh}{k}

\newcommand{\cR}{\mathcal{R}}

\newcommand{\dcR}{\delta\mathcal{R}}
\newcommand{\nx}{n_x}
\newcommand{\ny}{n_y}

\numberwithin{equation}{section}

\def\clock{{\count0=\time
           \divide\count0 60
           \ifnum\count0<10 0\fi\the\count0
           \multiply\count0 -60 \advance\count0 \time
           :\ifnum\count0<10 0\fi \the\count0
         }}
\newcommand{\timestamp}{{\small\vbox{\hbox{\tt \jobname.pdf}
\hbox{\the\day/\the\month/\the\year, \clock
}}}}

\begin{document}

\begin{titlepage}
\leftline{}
\vskip 2cm
\centerline{\LARGE \bf Rotating black holes and black bars at large $D$}
\vskip 1.2cm
\centerline{\bf Tom\'as Andrade$^{a}$, Roberto Emparan$^{a,b}$, David Licht$^{a}$}
\vskip 0.5cm
\centerline{\sl $^{a}$Departament de F{\'\i}sica Qu\`antica i Astrof\'{\i}sica, Institut de
Ci\`encies del Cosmos,}
\centerline{\sl  Universitat de
Barcelona, Mart\'{\i} i Franqu\`es 1, E-08028 Barcelona, Spain}
\smallskip
\centerline{\sl $^{b}$Instituci\'o Catalana de Recerca i Estudis
Avan\c cats (ICREA)}
\centerline{\sl Passeig Llu\'{\i}s Companys 23, E-08010 Barcelona, Spain}
\smallskip
\vskip 0.5cm
\centerline{\small\tt tandrade@icc.ub.edu, \, emparan@ub.edu,\, david.licht@icc.ub.edu}

\vskip 1.2cm
\centerline{\bf Abstract} \vskip 0.2cm 
\noindent 
We propose and demonstrate a new and efficient approach to investigate black hole dynamics in the limit of large number of dimensions $D$. The basic idea is that an asymptotically flat black brane evolving under the Gregory-Laflamme instability forms lumps that closely resemble a localized black hole. In this manner, the large-$D$ effective equations for extended black branes can be used to study localized black holes. We show that these equations have exact solutions for black-hole-like lumps on the brane, which correctly capture the main properties of Schwarzschild and Myers-Perry black holes at large $D$, including their slow quasinormal modes and the ultraspinning instabilities (axisymmetric or not) at large angular momenta. Furthermore, we obtain a novel class of rotating `black bar' solutions, which are stationary when $D\to\infty$, and are long-lived when $D$ is finite but large, since their gravitational wave emission is strongly suppressed. The leading large $D$ approximation reproduces to per-cent level accuracy previous numerical calculations of the bar-mode growth rate in $D=6,7$. 

\end{titlepage}
\pagestyle{empty}
\small
\normalsize
\newpage
\pagestyle{plain}
\setcounter{page}{1}

%



\section{Introduction}

Recent advances have shown that many problems in black hole physics simplify in an efficient way when expanded in the inverse of the number $D$ of spacetime dimensions \cite{Asnin:2007rw}--\cite{Kundu:2018dvx}. The studies of linearized perturbations in \cite{Emparan:2014aba} (greatly aided by the numerical work of \cite{Dias:2014eua}) revealed that the low frequency modes near the horizon of a black hole decouple, at all perturbation orders in $1/D$, from the faster oscillations that propagate away from the horizon. This has enabled the formulation of effective non-linear theories for the slow fluctuations of the black hole, which by now have been derived and put to use in several contexts.

In some instances the effective equations are more manageable than in others.
They are particularly simple for black branes, either asymptotically flat or asymptotically AdS \cite{Emparan:2015gva,Dandekar:2016jrp,Emparan:2016sjk}. In this article we focus on the equations for asymptotically flat black branes, which have been easily solved to study the main dynamical feature of these objects: the linearized instability discovered by Gregory and Laflamme (GL) \cite{Gregory:1993vy}, as well as its subsequent non-linear evolution.

Surprisingly, we have found that these equations are useful beyond their application to black branes: they also manage to capture a remarkable amount of the physics of localized black holes ---not only the known Schwarzschild and Myers-Perry (MP) variety \cite{Myers:1986un}, but also novel, unanticipated black holes at large $D$. In this article we initiate this approach with the study of individual neutral black holes, old and new, and their stability.

\bigskip
\noindent\textit{Black holes from bulging black branes}
\medskip

\begin{figure}[t]
\centerline{\includegraphics[width=.65\textwidth]{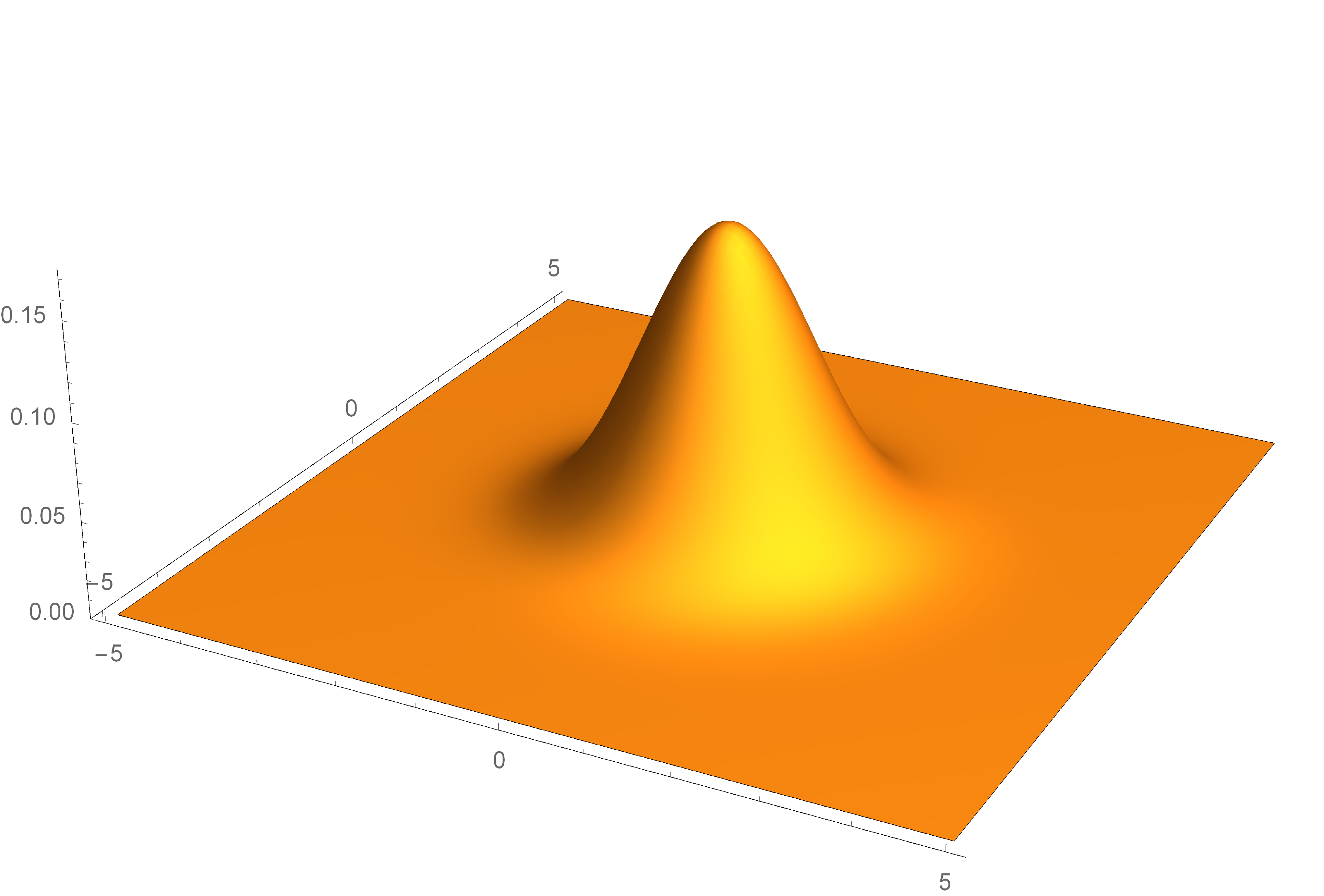}}
\caption{\small Black holes as gaussian lumps. Driven by the GL instability, a black membrane develops a bulge where most of its area and mass are concentrated with a profile that, when $D\to\infty$, is a gaussian, as shown in the figure. Representing Schwarzschild and Myers-Perry black holes in this manner accurately captures much of their physics, including their quasinormal vibrations. \label{fig:gaussian}}
\end{figure}

\noindent The basic idea can be easily understood. Consider the simplest case of a black string compactified on a long circle. Driven by the GL instability, the black string becomes increasingly inhomogeneous and develops lumps along its length. The endpoint of the instability depends on the number of spacetime dimensions \cite{Sorkin:2004qq,Lehner:2010pn,Harmark:2007md,Emparan:2018bmi}, but for a thin initial string and at large enough $D$, the system stabilizes on a configuration with large blobs that approach the shape of a Schwarzschild black hole, and which are connected by thin tubes of black string \cite{Emparan:2015gva}. For a black brane the end result is similar: black-hole-like bulges on a thin membrane (see fig.~\ref{fig:gaussian}). We will show that there is a simple, exact solution of the effective black brane equations that describes a bulge with many of the physical properties of a Schwarzschild black hole. It can be boosted to move at constant velocity like a black hole would, and it also generalizes to a solution that rotates around its axis like a MP black hole.

In the large-$D$ approximation, the bulge on the brane is a good approximation for only a `cap' of the Schwarzschild (or Myers-Perry) black hole horizon. It may then be surprising that, even though the angular extent of this cap is small, $\sim 1/\sqrt{D}$, it is nevertheless large enough to accurately account for much of the physics of the black hole when $D\gg 1$.

In order to illustrate how this is possible, let us take a sphere $S^D$, built as a fibration of spheres $S^{D-2}$ over a hemisphere (a topological disk) parametrized by $\theta\in [0,\pi/2]$, $\phi\in [0,2\pi)$, such that its metric is
\beq
d\Omega_D = d\theta^2+\sin^2\theta\,d\phi^2+\cos^2\theta\, d\Omega_{D-2}\,.
\eeq
The area of this sphere can be computed in an exact, recursive manner as
\beqa
\Omega_D &=&\Omega_{D-2}\,2\pi \int_{0}^{\pi/2} d\theta\,\sin\theta \lp\cos\theta\rp^{D-2}\nn\\
&=&\frac{2\pi}{D-1}\Omega_{D-2}\label{OmegaD}\,.
\eeqa
However, for us it is more interesting to observe that when $D$ is very large the above integrand is strongly peaked around the center of the hemisphere, $\theta\approx 0$. We can then estimate the integral using a saddle-point approximation. If we make
\beq\label{thinequ}
\theta= \frac{r}{\sqrt{D}}\,,
\eeq
so that $\cos\theta\approx 1-r^2/(2D)$, then the `density of $S^{D-2}$-area' on the plane $(r,\phi)$ becomes
\beq
a(r)=\lp\cos\theta\rp^{D-2}\approx \,e^{-r^2/2}\,,
\eeq
and we compute
\beq
\Omega_D\approx \frac{2\pi}{D}\Omega_{D-2}\int_{0}^{\infty} dr\,  r\,e^{-r^2/2}=\frac{2\pi}{D}\,\Omega_{D-2}\,,
\eeq
which is indeed the leading order approximation to the exact result \eqref{OmegaD} when $D\gg 1$. The upshot is that almost all of the area of the sphere is concentrated with a gaussian profile in a section of small angular extent $\Delta\theta\sim 1/\sqrt{D}$.

When the sphere is that of a large-$D$ black hole, essentially the same argument reveals that not only its area, but also the mass and other extensive quantities, as well as the far-zone gravitational potential $1/r^{D-3}$ of the black hole, can be recovered from a small cap, which can be alternatively viewed as a gaussian bulge on a black brane \cite{Suzuki:2015axa,Emparan:2015hwa}. More generally, if we view the $S^D$ as a fibration of $S^{D-p}$ over a ball $B_p$ (with $p\geq 1$ finite as $D\to\infty$), then the black hole is very well approximated by a gaussian bulge of width $\sim 1/\sqrt{D}$ on a black $p$-brane.

This observation had essentially been made already in \cite{Emparan:2015hwa} and \cite{Suzuki:2015axa}. What we have discovered, and will demonstrate in this article, is that this small cap of the horizon is sufficient to capture not only static, integrated magnitudes of the black hole: it also contains dynamical information such as the spectrum and profiles of its quasinormal excitations. Moreover, the simplicity of the equations has allowed us to identify a new kind of black hole, namely a rotating black bar, which is stationary when $D\to\infty$ but should be slowly radiating and long-lived for large but finite $D$. The instability of rotating black holes that is associated to this new branch of solutions is captured by our leading large-$D$ calculation at per-cent level accuracy when compared to earlier numerical results \cite{Shibata:2010wz,Dias:2014eua}.

Our study can be usefully related and compared with the effective theory of stationary black holes at large $D$ that was derived and solved in \cite{Suzuki:2015iha}. In that theory, the singly-spinning MP black hole is represented as an spheroidal elastic membrane rotating in a flat background. The theory also yields the quasinormal spectrum of the MP black hole. In our approach, the bulges of the membrane  match the spheroidal membrane around the symmetry axis on a region of angular size $\sim 1/\sqrt{D}$. The two descriptions complement each other well: the equations of \cite{Suzuki:2015iha} capture all of the horizon, but ours have finer resolution in the region near the symmetry axis, as we will argue. When the two approaches overlap we find perfect agreement; in particular, the quasinormal spectra coincide exactly.

Very recently, the elegant formulation of the large-$D$ effective theory by Bhattacharyya, Minwalla and collaborators \cite{Bhattacharyya:2015dva,Bhattacharyya:2015fdk} has been applied in \cite{Mandlik:2018wnw} to describe the simplest stationary black holes, namely, MP black holes and black rings, in good agreement with the construction of \cite{Suzuki:2015iha,Tanabe:2015hda}. It should be interesting to also investigate our approach within this framework.

\bigskip
\noindent\textit{Plan}
\medskip

\noindent In sec.~\ref{sec:effeqns} we lay down the approach based on the effective equations for black branes. In sec.~\ref{sec:statmaster} we present an analysis of stationary configurations that we will use later. In sec.~\ref{sec:eff21} we give the explicit form of the equations for $2+1$ membranes. These equations are then solved in sec.~\ref{sec:axisymm} to reproduce the axisymmetric Schwarzschild and Myers-Perry black holes at large $D$. In sec.~\ref{sec:bars} we obtain new non-axisymmetric rotating black bar solutions. In sec.~\ref{sec:qnms} we compute the quasinormal spectrum of Myers-Perry solutions, and in sec.~\ref{sec:pertbars} we find corotating zero modes of black bars. We conclude in sec.~\ref{sec:concl}.

The appendices discuss important issues that involve technicalities besides those in the main body of the article.
Appendix~\ref{app:SchwMP} shows in detail how the known Schwarzschild and Myers-Perry solutions become, in the limit $D\to\infty$, the same solutions for bulges on a membrane as we find in sec.~\ref{sec:axisymm}. Appendix~\ref{app:sphhar} shows that our perturbations of gaussian bulges match the spheroidal harmonic decomposition of perturbations of the MP black hole, and clarifies the relation to the study in \cite{Suzuki:2015iha}. Appendix~\ref{app:nonlin} discusses the behavior at large radius of the solutions to our equations, and issues of existence beyond linear perturbation theory.

\section{Effective black brane equations}\label{sec:effeqns}

We begin with an extensive review of the effective theory of large-$D$ black branes as derived in \cite{Emparan:2015gva} and further developed in \cite{Emparan:2016sjk}. Readers familiar with this material may skip this section, and then quickly skim over secs.~\ref{sec:statmaster} and \ref{sec:eff21} to arrive at the solutions in sec.~\ref{sec:axisymm}.

We write the metric for a dynamical, neutral, vacuum black $p$-brane to leading order at large-$D$ in Eddington-Finkelstein coordinates as
\beq\label{largeDmetric}
ds^2=2dtd\br -A dt^2-\frac2n C_i d\sigma^i dt +\frac1n G_{ij}d\sigma^i d\sigma^j +\br^2 d\Omega_{n+1}\,,
\eeq
with $n=D-3-p$. The radial coordinate orthogonal to the brane is $\rho$, and $\sigma^i$, $i=1,\dots, p$, are spatial coordinates along the brane. The lengths along these directions have been rescaled by a factor $1/\sqrt{n}$. The coordinate $t$ is null in \eqref{largeDmetric} but it will play the role of time in the effective membrane theory, which can be regarded as living on a surface at a distance in $\rho$ of order $1/n$ away from the horizon. Note also that the time scales we consider are order one in $n$ (\ie\ $\ord{n^0}$), and thus are slow compared to the short time scales $\ord{1/n}$ of the fast quasinormal frequencies that are integrated out in the effective theory.

The radial dependence in the Einstein equations can be solved to determine that the metric functions are
\beqa\label{ACG}
A&=&1-\frac{m(t,\sigma)}{\sR}\,,\\
C_i&=&\frac{p_i(t,\sigma)}{\sR}\,,\\
G_{ij}&=&\delta_{ij}+\frac1n \frac{p_i(t,\sigma)p_j(t,\sigma)}{m(t,\sigma)\sR}\,,
\eeqa
where we have introduced the near-horizon radial coordinate $\sR=\br^n$.
Then, the remaining Einstein equations reduce to a set of effective equations for the functions $m(t,\sigma)$ and $p_i(t,\sigma)$, namely 
\beqa
\partial_t m-\nabla^2 m&=&-\nabla_i p^i\,,\label{dyna}\\
\partial_t p_i-\nabla^2 p_i&=&\nabla_i m -\nabla_j\lp\frac{p_ip^j}{m}\rp\,,\label{dynb}
\eeqa
where the derivatives are taken in the flat spatial geometry $\delta_{ij}d\sigma^i d\sigma^j$.\footnote{The generalization to curved membrane geometries has been given in \cite{Andrade:2018zeb}.}

These equations encode the effective non-linear dynamics of a black brane at large $D$ and we will use them to investigate the properties of black holes in this limit. As we explained in the introduction, 
our approach is based on the fact that the uniform black branes (the solutions with $m=1$, $p_i=0$) are unstable and tend to localize into black-hole-like lumps.

We have written the effective equations as resembling diffusion equations, but they can also be recast in other forms. For instance, they take on a hydrodynamical aspect if instead of $p_i$ we use the velocity $v_i$, such that
\beq
p_i=\nabla_i m+m v_i\,.
\eeq
Then the equations \eqref{dyna}, \eqref{dynb} are those of mass continuity
\beq\label{masscont}
\pd_t m +\nabla_i(mv^i)=0\,,
\eeq
and momentum continuity
\beq\label{momcont}
\pd_t (m v_i)+\nabla^j\lp m v_i v_j +\tau_{ij}\rp=0\,,
\eeq
with effective stress tensor
\beq\label{effstress}
\tau_{ij}=- m\delta_{ij} -2m\nabla_{(i}v_{j)}-m\nabla_j\nabla_i\ln m\,.
\eeq

The function $m(t,\sigma)$ is (up to constant factors) the effective mass density of the black brane. Since the horizon is at $\sR=m$, we see that (to leading order in $1/D$) $m$ is also equal to the area density. The radius of the horizon is then
\beq
\rho_H=1+\frac{\cR(t,\sigma)}{n}+\ord{1/n^2}\,,
\eeq
where we have introduced the field
\beq\label{cR}
\cR(t,\sigma)=\ln m(t,\sigma)
\eeq
as a convenient measure of the area-radius. This radius variable is useful for another, more geometric interpretation of the effective equations and their solutions: the elastic `soap bubble' viewpoint discussed in sec.~\ref{subsec:elastic}. 

The effective equations have two important symmetries. First, if we perform a Galilean boost
\beq\label{boost}
\sigma_i \to \sigma_i -u_i t\,,
\eeq
then the velocity gets shifted accordingly,
\beq
v_i \to v_i +u_i\,.
\eeq
The symmetry is Galilean rather than relativistic since when $D$ is large the effective speed of sound on the black brane decreases as $\sim 1/\sqrt{D}$. In \eqref{largeDmetric} we rescale the lengths so as to maintain this speed finite.
 
Second, the equations are invariant under constant rescalings $m\to \lambda m$, $p_i\to\lambda p_i$. This symmetry corresponds to the scaling invariance of the vacuum Einstein equations, and allows to fix a reference scale arbitrarily, \eg\ to fix the mass of the solutions.

The spatial integrals of $m$, $p_i$, and $m v_i$ are conserved in time when spatially periodic boundary conditions are imposed, or when these fields vanish at infinity. These are the conservation laws of mass and momenta. In order to avoid awkward overall factors in the expressions for mass and angular momenta, we will work with the angular momentum (on the plane $(ij)$) per unit mass, which is obtained as
\beq
\frac{J_{ij}}{M}=\frac{\int d^p\sigma\, m (\sigma_i v_j-\sigma_j v_i)}{\int d^p\sigma\, m}=
\frac{\int d^p\sigma\, (\sigma_i p_j-\sigma_j p_i)}{\int d^p\sigma\, m}\,.
\eeq

\section{Stationary configurations}\label{sec:statmaster}

\subsection{Master equation}

We define stationary configurations as those that are invariant under the action of a vector
\beq
k=\pd_t+v^i\pd_i\,.
\eeq
Observe that we do not require that the solution be time-independent. Nevertheless, $k(m)=0$ determines that
\beq\label{dtm}
\pd_t m=-v^i\pd_i m\,.
\eeq
Then the mass continuity equation \eqref{masscont} becomes the condition that the expansion vanishes
\beq\label{zeroexp}
\nabla_i v^i=0\,.
\eeq
If in addition we require that $k$ is a Killing vector, $\nabla_{(\mu}k_{\nu)}=0$, then we obtain that the velocity flow is also time-independent and shear-free,
\beq
\pd_t v^i=0\,,\qquad \nabla_{(i}v_{j)}=0\,.
\eeq
It may not be obvious that one should require that $k$ be a Killing vector, but this can be proved from the requirement of absence of shear and expansion \cite{Caldarelli:2008mv}, since otherwise viscosity (both shear and bulk) would generate dissipation in the system. The velocity flow may still have vorticity. 

Under these conditions we can follow the steps in \cite{Emparan:2016sjk} and show that, using the membrane radius $\cR$ \eqref{cR}, we can write
\beq
\partial_t(m v_i)+\nabla^j(m v_i v_j)=-\frac12 e^\cR \nabla_i (v^2)
\eeq
and
\beq
\nabla^j\tau_{ij}=-e^\cR \nabla_i \lp \cR +\nabla^2 \cR +\frac12(\nabla\cR)^2\rp\,.
\eeq
This reduces the momentum continuity equations \eqref{momcont} to a single equation for  $\cR$,
\beq\label{stateq}
\nabla^2\cR+\frac12 (\nabla \cR)^2+\cR +\frac{v^2}2 =0\,,
\eeq
where we have absorbed an integration constant by appropriately shifting the value of $\cR$.

Eq.~\eqref{stateq} is the master equation that governs the stationary sector of solutions. The derivation of this equation in \cite{Emparan:2016sjk} made the assumption that $\pd_t m=0$. We now see that it also applies to more general stationary configurations in which \eqref{dtm} holds instead.

\subsection{Elastic viewpoint}\label{subsec:elastic}

The master equation \eqref{stateq} admits an interpretation as an elasticity equation for a soap bubble \cite{Emparan:2015hwa,Suzuki:2015iha,Emparan:2016sjk}. To see this, consider embedding the surface
\beq
\rho=1+\frac{\cR(\sigma)}{n}
\eeq
in a constant-time section of the Minkowski geometry that appears at large $\sR$ in \eqref{largeDmetric}, \eqref{ACG},
\beq\label{backmink}
ds^2=-d\hat{t}^2+d\rho^2 +\frac1{n}\delta_{ij}d\sigma^i d\sigma^j +\rho^2 d\Omega_{n+1}\,
\eeq
(we have changed from the Eddington-Finkelstein null coordinate $t$ in \eqref{largeDmetric} to Minkowski time $\hat{t}$). The trace of the extrinsic curvature of this surface is
\beq\label{Kbrane}
K=n+1-\lp \cR +\nabla^2 \cR +\frac12(\nabla\cR)^2\rp+\ord{1/n}\,.
\eeq
Then eq.~\eqref{stateq} can be written, up to next-to-leading order in $1/n$, as
\beq\label{soap}
\sqrt{1-\mathbf{v}^2}K=\textrm{constant}\,,
\eeq
where $\mathbf{v}=v/\sqrt{n}$ is the physical velocity along the brane, since the lengths in \eqref{backmink} are rescaled by a factor $1/\sqrt{n}$. The Young-Laplace equation $K=\mathrm{constant}$ famously describes the shape of soap bubbles (more generally, interfaces between fluids). Eq.~\eqref{soap}, which was first derived for large-$D$ black holes in \cite{Suzuki:2015iha}, includes a Lorentz-redshift factor that accounts for the possible rotation of the bubble (or any motion along its surface).

This elastic interpretation of the effective theory allows to make sense of some features of the effective equations that remain obscure in the hydrodynamic version. In the latter, the constitutive relations \eqref{effstress} contain only one term at next-to-viscous order, and none at higher gradient order. But the large-$D$ expansion is not an expansion in worldvolume gradients. Why should the effective theory involve only a finite number of them?
The mystery dissipates (at least for stationary solutions) in the elastic interpretation, which leaves no room for any other form of $\tau_{ij}$ than precisely \eqref{effstress}, since this is the one that completes the expression for $K$ in \eqref{Kbrane}.

Furthermore, the elastic viewpoint, in which a membrane with positive tension forms soap bubbles, may look more natural than a hydrodynamic view where an unstable fluid with negative pressure (as in \eqref{effstress}) clumps into blobs of fluid.\footnote{For a black brane in AdS the hydrodynamic interpretation is apt since the large-$D$ effective fluid has positive pressure and is stable. Curiously, the effective elastic equation is one with positive tension, but the membrane is subject to a gravitational potential from the AdS geometry.},\footnote{Note the contrast with the blackfold approach of \cite{Emparan:2009at} (which is another effective theory of black brane dynamics): in the latter some collective degrees of freedom are elastic and others are hydrodynamic, whereas in the large-$D$ effective theory the same degrees of freedom admit one or the other interpretation.}

\section{Effective $2+1$ membrane equations}\label{sec:eff21}

Henceforth we will consider membranes (2-branes), for which we write the spatial geometry as
\beq
ds^2=dr^2+r^2d\phi^2
\eeq
and set
\beq
p=p_r dr +p_\phi d\phi\,.
\eeq
The equations take the form
\beq\label{eqmpolar}
\partial_t m=\pd^2_r m+\frac{\pd_r m}r+\frac{\pd^2_\phi m}{r^2}-\pd_r p_r-\frac{p_r}r -\frac{\pd_\phi p_\phi}{r^2}\,,
\eeq
\beqa\label{eqprpolar}
\pd_t p_r&=&\pd_r m+\pd_r^2 p_r +\frac{\pd_r p_r}r-\frac{p_r}{r^2}+\frac{\pd_\phi^2 p_r}{r^2}-\frac{2\pd_\phi p_\phi}{r^3}\nn\\
&&-\pd_r\lp\frac{p_r^2}{m}\rp -\frac{p_r^2}{r m}-\frac1{r^2}\pd_\phi\lp\frac{p_r p_\phi}{m}\rp+\frac{p_\phi^2}{r^3 m}\,,
\eeqa
\beqa\label{eqpthpolar}
\pd_t p_\phi &=& \pd_\phi m+\frac{\pd_\phi^2 p_\phi}{r^2}-\frac{\pd_r p_\phi}{r}+\pd_r^2 p_\phi+\frac2r\pd_\phi p_r\nn\\
&& -\frac1{r^2}\pd^2_\phi \lp\frac{p_\phi^2}{m}\rp-\pd_r\lp\frac{p_r p_\phi}{m}\rp-\frac{p_r p_\phi}{r m}\,.
\eeqa

\subsection{Time-independent axisymmetric configurations}

For configurations that are independent of $t$ and $\phi$, eq.~\eqref{eqmpolar} is solved by
\beq
p_r=\pd_r m\,,
\eeq
\ie $v_r=0$.
Using this, eq.~\eqref{eqpthpolar} can be rewritten as
\beq
\pd_r \lp m r^3 \pd_r \frac{p^\phi}{m}\rp =0\,,
\eeq
where $p^\phi=g^{\phi\phi}p_\phi=p_\phi/r^2$. If we require that $m$ and $p^\phi$ asymptote to finite values at $r\to\infty$, then we can integrate this equation to conclude that
\beq
\frac{p^\phi}{m}=\Omega
\eeq
is a constant, equal to the angular velocity of the horizon. That is,
\beq
p_\phi =m r^2  \Omega\,.
\eeq
Hence we have proven that the rotation velocity $v^\phi =\Omega$ is a constant, which in sec.~\ref{sec:statmaster} we had assumed. Thus we have given the large-$D$ proof of the black hole rigidity theorem.

The remaining equation \eqref{eqprpolar}, in terms of $\cR(r)=\ln m(r)$, is indeed a particular case of the master stationary equation \eqref{stateq}, namely,
\beq\label{stateqpol}
\cR''+\frac{\cR'}{r}+\frac12 {\cR'}^2+\cR+\frac{\Omega^2r^2}2=0\,.
\eeq

\subsection{Stationary membrane master equation}

Consider a configuration that is stationary but not necessarily axisymmetric nor time-independent, and take a purely rotational velocity field,
\beq\label{rotflow}
v^\phi=\Omega\,.
\eeq
In this case the master equation \eqref{stateq} for 
\beq
\cR=\cR(r,\phi-\Omega t)
\eeq
takes the form
\beq\label{stateq2}
\pd^2_r \cR+\frac{\pd_r \cR}{r}+\frac{\pd_\phi^2 \cR}{r^2}+\frac12 \lp \lp\pd_r \cR\rp^2+\frac{\lp \pd_\phi \cR\rp^2}{r^2}\rp +\cR +\frac{\Omega^2 r^2}{2}=0\,.
\eeq
In our subsequent analysis we will make extensive use of this equation and of \eqref{stateqpol}, to which \eqref{stateq2} reduces for axisymmetric profiles.

The mass density is
\beq
m(r,\phi-\Omega t)=e^{\cR(r,\phi-\Omega t)}
\eeq
and the $p_i$ are
\beqa\label{pistat}
p_r&=&\pd_r m\,,\\
p_\phi&=&\pd_\phi m+\Omega\, r^2 m\,.
\eeqa

\section{Axisymmetric black holes}\label{sec:axisymm}

In ref.~\cite{Emparan:2015gva} the numerical evolution of thin unstable black branes was found to stabilize at very approximately gaussian profiles  $m(r)\sim e^{-r^2/2}$. It was noticed in \cite{Emparan:2015hwa,Suzuki:2015axa} that these reproduce well the features of a Schwarzschild black hole. This is the main intuition that leads us to seek new solutions to the effective equations \eqref{dyna}, \eqref{dynb} that capture the physical properties of localized black holes ---even if, as we will see later, some of these are not always stable themselves.

When looking for stationary axisymmetric configurations we only need to solve the master equation \eqref{stateqpol}.
The rotation term $\sim \Omega^2 r^2$ suggests that we try an ansatz where $\cR$ is quadratic in $r$. This yields easily the solution
\beq\label{RMP}
\cR(r)=\cR_0-\frac{r^2}{2(1+a^2)}\,,
\eeq
with
\beq
\Omega=\frac{a}{1+a^2}\,.
\eeq
The constant $\cR_0$ can be chosen arbitrarily, but our specific choices in \eqref{stateqpol} require that $\cR_0=2/(1+a^2)$, and then
\beq\label{RMP2}
\cR(r)=\frac{2}{1+a^2}\lp 1-\frac{r^2}{4}\rp\,.
\eeq

The solution for the mass and area density has gaussian profile
\beq\label{mMP}
m(r)=m_0\exp\lp-\frac{r^2}{2(1+a^2)}\rp\,,
\eeq
with $m_0=e^{\cR_0}$, and
\beq\label{prMP}
p_r=\pd_r m =-m_0\frac{r}{1+a^2}\exp\lp-\frac{r^2}{2(1+a^2)}\rp\,,
\eeq
\beq\label{pphMP}
p_\phi=m r^2\Omega=m_0\frac{a r^2}{1+a^2}\exp\lp-\frac{r^2}{2(1+a^2)}\rp\,.
\eeq

The angular momentum per unit mass is
\beq
\frac{J}{M}=\frac{ \int_0^\infty dr\, r\, p_\phi(r)}{ \int_0^\infty dr\, r\, m(r)}=2 a\,.
\eeq
This reproduces correctly the large-$D$ value for MP black holes \cite{Emparan:2013moa}, once we take into account that lengths on the membrane are rescaled by a factor of $1/\sqrt{D}$. so the physical value is actually $J/(MD)=2a/D$.

\begin{figure}[t]
\centerline{\includegraphics[width=\textwidth]{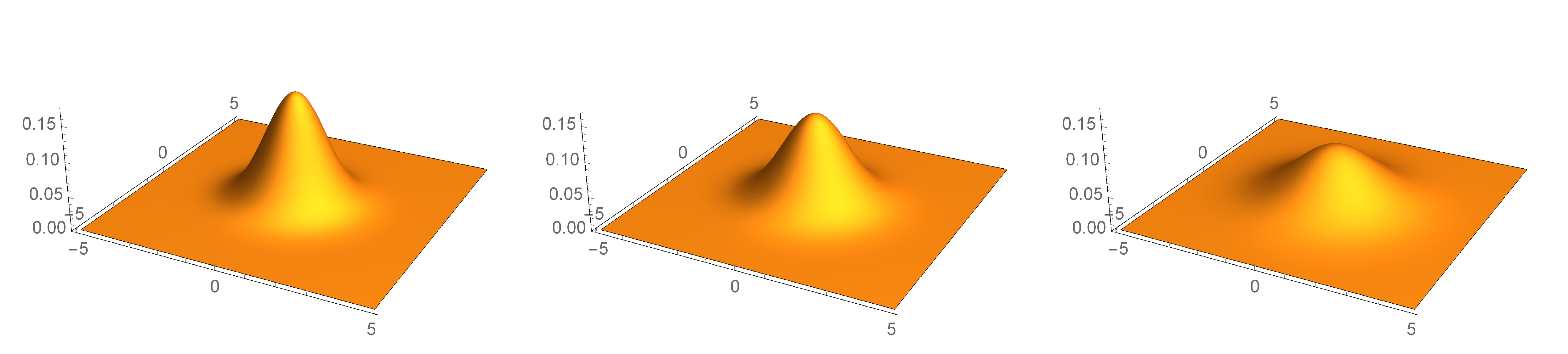}}
\caption{\small Mass density profiles $m(r)$ for black holes with $J/M=0$ (left), $J/M=1$ (middle), $J/M=2$ (right). \label{fig:spin012}}
\end{figure}

Note that $\Omega\in [0,1/2]$. The maximum $\Omega=1/2$ is achieved for $a=1$. On the other hand, the value $\Omega\to 0$ is reached both when $a\to 0$ and when $a\to\infty$. The former is the static limit, whereas the latter is the ultraspinning limit where $J/M\to\infty$. If we keep the mass fixed and increase the spin, the brane profile $m(r)$ flattens as it spreads over an area $\propto a^2$, see fig.~\ref{fig:spin012}.
The rotation parameter in terms of $\Omega$ is 
\beq
a=\frac{1\pm\sqrt{1-4\Omega^2}}{2\Omega}\,,
\eeq
and the $+$ sign gives the ultraspinning branch. 

In appendix~\ref{app:SchwMP} we show that these solutions can be recovered if we start with the known Schwarzschild-Tangherlini and Myers-Perry black holes, and take their limit when $D\to\infty$ in an appropriate manner. Thus, in the following we will refer to the solution \eqref{RMP2}, \eqref{mMP} as the MP black hole.

\subsection{Travelling rotating black holes}

The boost symmetry \eqref{boost} can be used to generate solutions that travel at constant speed $u_i$. Changing to cartesian coordinates $(x^1,x^2)$, the travelling rotating black hole is described by
\beq
m(t,x)=m_0\exp\lp -\frac{(x_i-u_i t)(x^i-u^i t)}{2(1+a^2)}\rp\,,
\eeq
\beq
v_i = u_i+\frac{a}{1+a^2}\varepsilon_{ij}(x^j - u^j t)\,.
\eeq

\section{Rotating black bars}\label{sec:bars}

Remarkably, it is also possible to find explicitly a class of stationary but time-dependent, non-axisymmetric exact solutions. 

\subsection{Solution}

Assume again that $\cR$ depends quadratically on $r$, but now allow an angle-dependent stationary profile\footnote{Regularity at the rotation axis $r=0$ requires that $\cR_0$ be $\phi$-independent.}
\beq
\cR=\cR_0-r^2 F(\phi-\Omega t)\,.
\eeq
Plugging this ansatz into \eqref{stateq2}, and setting for convenience $\cR_0=1$, we get an $r$-dependent equation for $F$. If we consider it at $r=0$ we get the equation
\beq
F''+4F-1=0\,,
\eeq
which, up to an arbitrary initial value of the phase, is solved by
\beq
F=\frac{1}{4}+C \cos\lp 2(\phi-\Omega t)\rp
\eeq
with constant $C$. Inserting this again in eq.~\eqref{stateq2}, but considering $r\neq 0$, we find an algebraic equation that is solved for
\beq
C=\pm \frac{\sqrt{1-4\Omega^2}}{4}\,.
\eeq
The sign choice here can be absorbed by a phase shift, so the solution we finally obtain is
\beq\label{Rbar}
\cR(t,r,\phi)=1-\frac{r^2}{4}\lp 1+\sqrt{1-4\Omega^2}\,\cos\lp 2(\phi-\Omega t)\rp\rp\,.
\eeq

The corresponding mass density is 
\beqa\label{mbar}
m(t,r,\phi)&=&\exp\lp 1-\frac{r^2}{4}\lp 1+\sqrt{1-4\Omega^2}\,\cos\lp 2(\phi-\Omega t)\rp\rp\rp\,,
\eeqa
and the $p_i$ are given by \eqref{pistat}.

This mass profile \eqref{mbar} has a dipolar dependence on the angle, and thus can be regarded as describing a `rotating bar', which extends along the angular directions
\beq
\phi=\Omega t \pm \frac{\pi}{2}\,.
\eeq
\begin{figure}[t]
\centerline{\includegraphics[width=.7\textwidth]{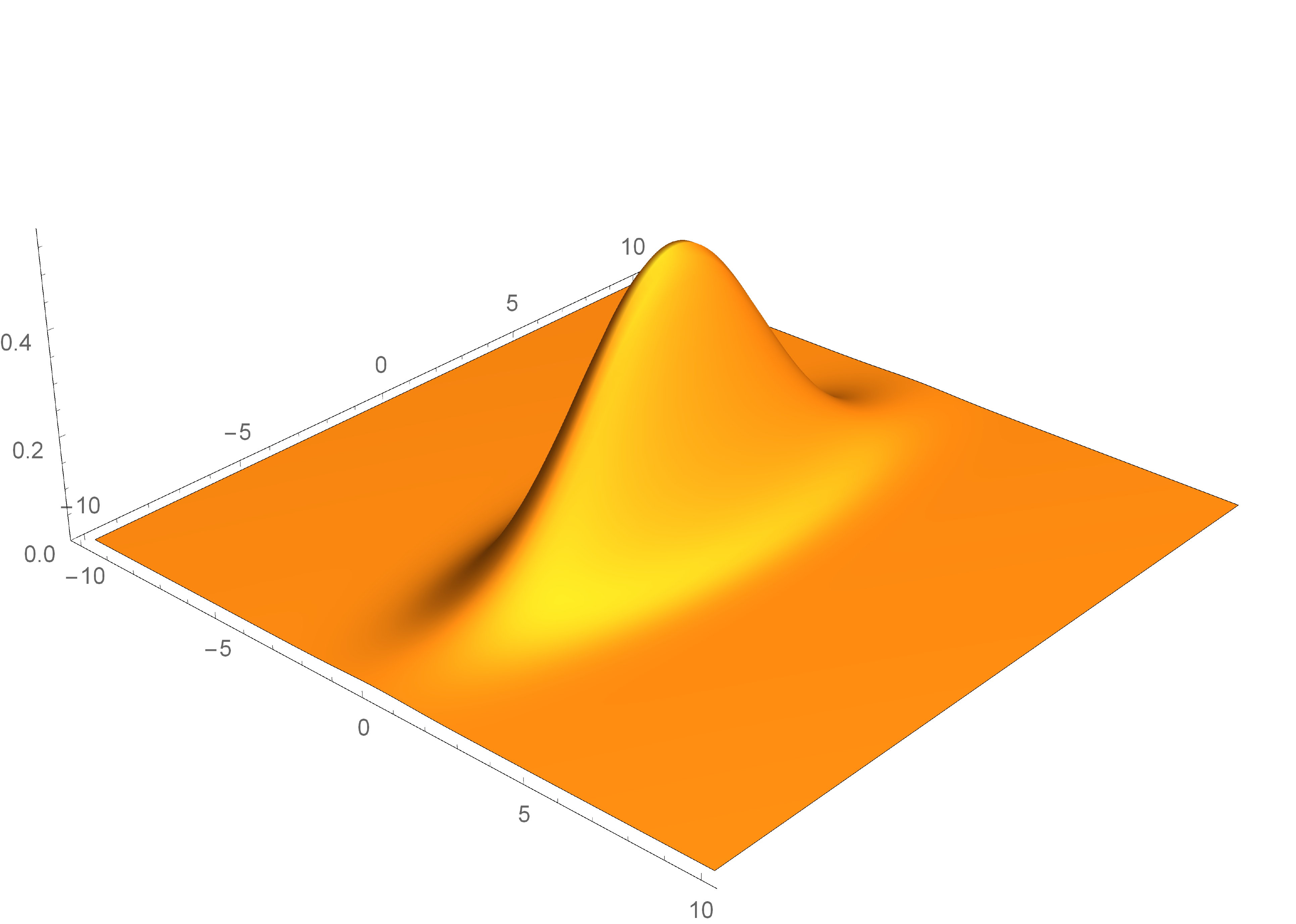}}
\caption{\small Mass density profile $m$ for the black bar with $\Omega=0.3$. \label{fig:bar}}
\end{figure}
It is also useful to present the solution in rotating Cartesian coordinates
\beqa
x&=&x^1 \cos\Omega t + x^2 \sin\Omega t=r\cos(\phi -\Omega t)\,,\nn\\
y&=&x^2 \cos\Omega t - x^2\sin\Omega t= r\sin(\phi -\Omega t)\,,\label{cartcorot}
\eeqa
such that $x$ and $y$ are corotating directions transverse to the bar and along the bar, respectively. If we introduce the length $\ell_\|$ and width $\ell_\perp$  of the bar,
\beqa
\ell_\|^2&=& \frac{2}{1-\sqrt{1-4\Omega^2}}\,,\nn\\
\ell_\perp^2&=& \frac{2}{1+\sqrt{1-4\Omega^2}}\,,\label{barells}
\eeqa
then the solution \eqref{Rbar} reads
\beq
\cR(x,y)=1-\frac{x^2}{ 2\ell_\perp^2}-\frac{y^2}{2\ell_\|^2}\,.
\eeq

\subsection{Physical properties}

The spin per unit mass of this solution is\footnote{Henceforth we assume $\Omega\geq 0$ without loss of generality.}
\beq
\frac{J}{M}=\frac1{\Omega}\,.
\eeq

The angular velocity in \eqref{Rbar} is restricted to $0\leq \Omega\leq 1/2$. When $\Omega=1/2$ we recover a rotating MP black hole \eqref{RMP} with $a=1$. Starting from this solution and decreasing $\Omega$ the profile \eqref{mbar} develops an increasingly longer and narrower shape (see fig.~\ref{fig:bar}), with longitudinal extent $\ell_\|$ and transverse thickness $\ell_\perp$. If we keep the mass fixed, the height $m_0$ of the bar decreases in proportion to $\Omega$.
In the limit $\Omega\to 0$ we have
\beq
\ell_\|\to \infty\,,\qquad\ell_\perp\to 1\,,
\eeq
and thus we recover an infinite, static black string.

\begin{figure}[t]
\centerline{\includegraphics[width=.75\textwidth]{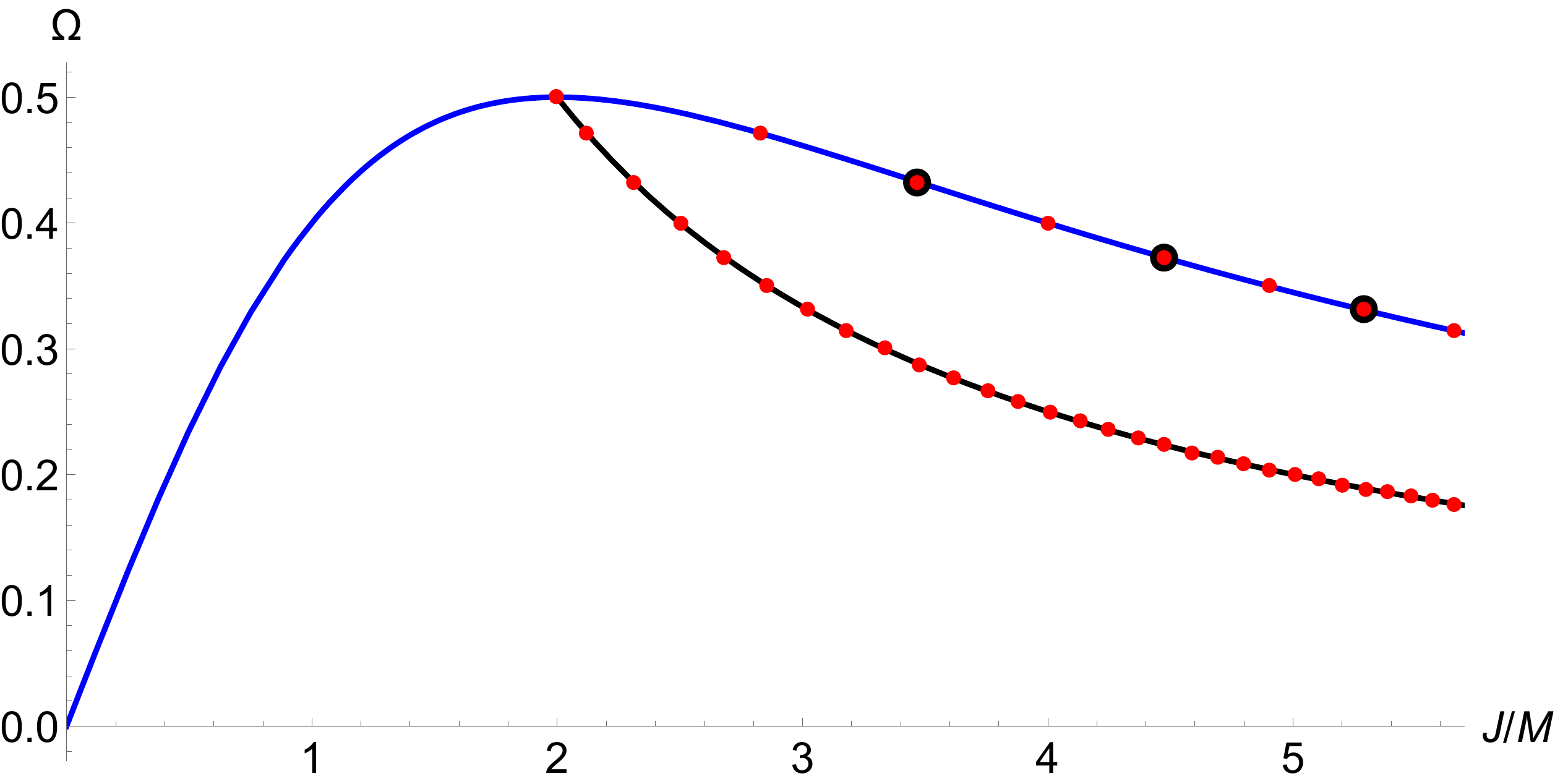}}
\caption{\small Phases of Myers-Perry black holes (blue line) and black bars (black line) in the plane $(J/M,\Omega)$. The red dots indicate the presence of non-axisymmetric corotating zero modes for MP black holes \eqref{MPnaxizero} and for black bars \eqref{goodmodes}. The ones in the MP branch encircled by a black dot indicate also axisymmetric zero modes \eqref{MPzero}. New families of solutions are expected to branch off all these points.\label{fig:OmJ}}
\end{figure}

At $\Omega=1/2$ we have a bifurcation into MP black holes and black bars, see fig.~\ref{fig:OmJ}.
In both the MP black hole and the black bar the angular velocity decreases from that point on, but for the same spin and mass, the bar rotates more slowly than the MP black hole. That is, it has a larger moment of inertia. The solutions with the same mass also have the same area, so we cannot predict on thermodynamic grounds which of the two solutions is preferred for given $M$ and $J$. This requires a computation to a higher order in the $1/D$ expansion. Nevertheless, since as we will see the MP black holes become unstable for $J/M>2$, and the black bars resemble black strings of finite length, we expect that the black bar is dynamically stable after the bifurcation\footnote{But only until the appearance of an instability of the black bar at $J/M=3/\sqrt{2}\approx 2.12$, see sec.~\ref{sec:pertbars}.} and also thermodynamically preferred over the MP black hole.

When $\Omega$ is small the length of the bar is $\ell_\|\sim 1/\Omega$ and we can easily see that $J$ and $M$ behave like in a rigidly rotating solid,
\beq
J\sim M \ell_\|^2 \Omega\,.
\eeq
A similar relation has been known since long ago to also hold for the ultraspinning MP black hole,  with $\ell_\|\sim a$ \cite{Emparan:2003sy} and also for black rings.

\subsection{Stationarity of black bars}\label{subsec:gws}

It may seem surprising that a rotating black bar exists as a stationary configuration: should it not be radiating gravitational waves?

The reason it does not is that in the large-$D$ limit the gravitational radiation decouples from the effective membrane \cite{Emparan:2014aba,Emparan:2015hwa,Bhattacharyya:2016nhn}. The decoupling actually holds to all perturbative orders in the $1/D$ expansion.\footnote{The rotating black bar would cease to be stationary, without still radiating, if the effective theory at higher perturbation orders in $1/D$ gave rise to dissipation from squared vorticity terms.} In other words, the decay rate of a black bar that is radiating gravitational waves must be non-perturbatively small in $1/D$.

We can easily estimate this rate. The bar has length $\sim\Omega^{-1}$, and when it rotates it emits waves with frequency $\Omega$. The radiating power is then
\beq
P\sim GM^2\Omega^{D-2}\,.
\eeq
Since $\Omega<1$, this implies that the decay rate
\beq
\frac{\dot{M}}{M}=\frac{P}{M}\sim GM\,\Omega^{D-2}
\eeq
is small exponentially in $D$.

Bar deformations of rotating black holes in $D=6,7$ have been observed in full numerical simulations in \cite{Shibata:2010wz}. These black bars spin down to a stable configuration through the emission of gravitational waves. The relaxation timescale in $D=6$ is found to be $\sim 100 r_0$, and in $D=7$ even larger than this. So indeed the black bars are long-lived, in agreement with our arguments.

\section{Quasinormal modes and stability of MP black holes}\label{sec:qnms}

Now we analyze the small, linearized perturbations of the previous solutions.

We begin by studying zero modes of the MP black holes, which leave the solution stationary. Their computation is simpler than the generic non-zero frequency quasinormal modes, and they are particularly important since they indicate the onset of instabilities and new branches of stationary solutions. When we solve afterwards for the modes with finite frequency, we will recover the zero modes as special cases.

\subsection{Corotating zero modes}
\label{sec:corotating}

We perturb the rotating MP solution \eqref{RMP2}, while remaining in the stationary sector, by making
\beq
\cR(r)=\frac{2}{1+a^2}\lp 1-\frac{r^2}{4}\rp+\epsilon\,\delta \cR(r) e^{i\nh(\phi-\Omega t)}\,,
\eeq
with
\beq
\Omega =\frac{a}{1+a^2}\,,
\eeq
and integer $\nh$. When $\nh=0$ these are axisymmetric, time-independent perturbations, and therefore are properly zero modes. When $\nh\neq 0$ their frequency
\beq\label{corotomega}
\omega=\nh\Omega
\eeq
is real and such that the perturbation corotates with the unperturbed black hole. Therefore we also regard them as zero modes. We can still use eq.~\eqref{stateq2} to study them. Note that corotating modes satisfying \eqref{corotomega} are at the threshold for superradiance. Below we will comment further on this point.

We adjust $\delta \cR$ by an additive constant so as to maintain $\cR_0$ fixed. Then we find the equation
\beq\label{coroteq}
\delta\cR''+\lp\frac1r-\frac{r}{1+a^2}\rp \delta\cR'+\lp 1-\frac{\nh^2}{r^2}\rp \delta\cR=0\,.
\eeq
As explained in appendix~\ref{app:sphhar}, this is precisely the equation for spheroidal harmonics on $S^{D-2}$, in the limit $n\to\infty$, when we focus on small polar angles $\theta\sim r/\sqrt{n}$. The appearance of this equation in this problem is a remarkable illustration of the extent to which the black brane equations \eqref{dyna}, \eqref{dynb}, manage to capture the dynamics of localized black holes. Even though eqns.~\eqref{dyna} and \eqref{dynb} should seemingly only know about fluctuations of planar black branes, they also describe accurately the vibrations of a spheroidal, large-$D$ MP black hole. This explains why in our subsequent analyses we will repeatedly encounter this equation.

Eq.~\eqref{coroteq}  has a regular singular point at $r=0$ and an irregular point at $r=\infty$, and it can be transformed into a confluent hypergeometric equation. 
The solutions that are regular at $r=0$ and which avoid non-analytic behavior $\sim e^{r^2}$ at $r\to\infty$ are expressed in terms of associated Laguerre polynomials $L_k^{|\nh|} (x)$, in the form
\beq\label{cRLaguerre}
\delta\cR(r)= r^{|\nh|} L_k^{|\nh|}\lp \frac{r^2}{2(1+a^2)}\rp\,, 
\eeq
with non-negative integer index
\beq
k=\frac{a^2+1-|\nh|}2\,.
\eeq
Then, MP black holes admit corotating zero mode perturbations only when the rotation parameter has the critical value
\beq\label{azero}
a_c^2=|\nh|+2k-1\,,\qquad k=0,1,2,\dots
\eeq
The index $k$ has the interpretation of a `radial overtone' number, such that, for a given value of $\nh$, the number of oscillations along $r$ increases with $k$. 
It is convenient to introduce the angular momentum number $\ell$
\beq\label{ellkm}
\ell=2k+|\nh|
\eeq
for the spherical harmonics of $S^{D-2}$ (see appendix~\ref{app:sphhar}), in terms of which the critical values of the rotation are
\beq\label{acrit}
a_c^2=\ell-1\,.
\eeq
These values are the same as found in \cite{Suzuki:2015iha}.  The behavior of \eqref{cRLaguerre} at large $r$, $\cR\sim r^\ell$, also matches the dependence $\sim \theta^\ell$ at small $\theta$ of the mode solutions in \cite{Suzuki:2015iha}.

We analyze now these solutions in more detail. Note that the solution for $|\nh|=1$ and $k=0$ corresponds to a shift of the center of the Schwarzschild solution away from $r=0$, so it is pure gauge.

\paragraph{Axisymmetric, time-independent zero modes.}

These are obtained when $\nh=0$. In this case the solutions are Laguerre polynomials,
\beq\label{axizeroL}
\delta\cR(r)=L_k\lp \frac{r^2}{4k}\rp\,,
\eeq
for $k=1,2,\dots$

When $k=1$ and therefore $a=1$, we have
\beq
\delta\cR(r)=1-\frac{r^2}{4}\,,
\eeq
which does not yield any new solution: it is a perturbation that varies $a$, adding angular momentum to the MP black hole while remaining in the same family of solutions. It has been known for some time that these zero mode deformations of MP black holes exist at the maximum of $\Omega$ in any $D\geq 6$ \cite{Dias:2009iu}.

The modes for $k=2,3,\dots$, which appear at 
\beq\label{MPzero}
a=\sqrt{3},\,\sqrt{5},\,\sqrt{7},\dots,\qquad \Omega=\frac{\sqrt{3}}{4}\,,\frac{\sqrt{5}}{6}\,,\frac{\sqrt{6}}{7}\,,\dots
\eeq
are deformations that should lead to new branches of stationary axisymmetric `bumpy black holes'. These had been first conjectured to exist in $D\geq 6$ in \cite{Emparan:2003sy}. The zero modes were explicitly constructed numerically in $D=6,7,8$ in \cite{Dias:2009iu}, and their non-linear extension in \cite{Dias:2014cia,Emparan:2014pra}. 

\paragraph{Non-axisymmetric, corotating zero modes.}

When we consider $|\nh|>0$ there are zero modes for
\beq\label{MPnaxizero}
a=1,\,\sqrt{2},\,\sqrt{3},\,2,\dots,\qquad \Omega=\frac12\,,\frac{\sqrt{2}}{3}\,,\frac{\sqrt{3}}{4}\,,\frac{2}{5}\,,\dots
\eeq
Observe that when different values of $\nh$ and $k$ combine to give the same value of $|\nh|+2k$, then the same solution admits several zero modes. 

The `fundamental' modes with $k=0$ have a simple radial profile,
\beq
\delta\cR(r) =r^{|\nh|}\,.
\eeq
Of these modes, the first non-trivial one, with $|\nh|=2$, corresponds to the black bar in a perturbative expansion around $a=1$, namely (taking a real perturbation)
\beq\label{blackbarmode}
\cR(r)= 1-\frac{r^2}{4}+\epsilon\,r^2 \cos(2\phi-t)\,.
\eeq
Higher values of $|\nh|$ signal new branches of solutions with higher multipole-bar deformations. 

Solutions with $k>0$ involve additional, higher powers of $r$, so they have more radial oscillations and can be regarded as `bumpy $\nh$-bar modes'.

\subsection{Quasinormal modes}

In order to study the spectrum of QNMs at arbitrary finite frequency, we need to turn to the 
full equations \eqref{eqmpolar} \eqref{eqprpolar}, \eqref{eqpthpolar}. We perturb
around the MP black hole, taking
\begin{align}
	m &= \bar m(r) + \epsilon\, e^{- i \omega t + i \nh \phi}\, \delta m(r), \\
	p_r &= \bar p_r (r) + \epsilon\, e^{- i \omega t + i \nh \phi}\,  \delta p_r(r), \\
	p_\phi &= \bar p_\phi (r) + \epsilon\, e^{- i \omega t + i \nh \phi}\,  \delta p_\phi(r),
\end{align}
\noindent with the background $\bar m$, $\bar p_r$, $\bar p_\phi$ given by 
\eqref{mMP}, \eqref{prMP}, \eqref{pphMP}, and working to linear order in $\epsilon$. 
By doing so, we obtain three coupled ODEs for the radial profiles. 
Quite remarkably, the spectrum of QNM can be found analytically. To do so, we first find a 
decoupled sixth order equation for $\delta m$, which can be easily obtained by taking linear combinations 
of the fluctuation equations. This equation takes the form 
\begin{equation}
	\mathcal{L}_1\mathcal{L}_2 \mathcal{L}_3 \delta \cR(r) = 0 \,,
\end{equation}
where (up to irrelevant constant factors)
\begin{equation}\label{delR}
	\delta \cR (r)=\frac{\delta m(r)}{\bar{m}(r)}= \exp\lp \frac{r^2}{2(1+a^2)}\rp \delta m(r)\,,
\end{equation}
\noindent and $\mathcal{L}_i$ are three second order, linear differential operators of the form 
\begin{equation}
\label{def Li}
	\mathcal{L}_i  = \frac{d^2}{dr^2} + \left( \frac{1}{r} - \frac{r}{1+a^2} \right)\frac{d}{dr} +  \frac{\nh + 2 k_i}{1+a^2}  - \frac{\nh^2}{r^2} \,.
\end{equation}
Here we recognize again the confluent hypergeometric operators of appendix~\ref{app:sphhar}, whose eigenfunctions are spheroidal harmonics of the $S^{D-2}$ at large $D$.

The constants $k_i$ are the three roots of the cubic equation
\beqa\label{cubick}
0&=&k^3-\frac{\gamma_\omega}2 k^2+\frac{\gamma _{\omega } \left(1-a^2+\gamma _{\omega }\right)+2 \left(a^2-3\right)-3 (a-i)^2 \nh }{12}k\\
&+&\frac{ \left(a^2+6 i a-\gamma _{\omega }-3\right) \left(a^4-6 i a^3-2 (a-3 i) a \gamma _{\omega }-9 (a-i)^2 \nh-18 i a+\gamma _{\omega }^2-9\right)}{216}\,,\nn
\eeqa
where we have introduced
\begin{equation}
	\gamma_\omega = 3 i \left(a^2+1\right) \omega +a^2-3 i a \nh-3 \nh+3\,.
\end{equation}

For any $k_i$ the operators \eqref{def Li} commute. Thus, the profiles correspond to the solutions of 
$\mathcal{L}_i \delta \cR  =0$, for $i= 1,2,3$. Since the $k_i$ are all roots of the same polynomial, the three equations are equivalent. The solutions that are regular at $r=0$ and at infinity are given by the associated Laguerre polynomials \eqref{cRLaguerre} with
\beq
k=0,1,2,3\dots
\eeq
This imposes a quantization condition on the frequencies that appear in \eqref{cubick}, which is itself a cubic equation in $\omega$. In order to write it more manifestly as such, it is convenient to use, instead of $k$, the angular momentum parameter $\ell$ of \eqref{ellkm}. Using it, \eqref{cubick} can be rewritten as\footnote{This agrees with 
the result of \cite{Suzuki:2015iha}, after correcting typos in their eq.~(4.20).}
\beqa
\label{w cubic quant}
0&= & \omega ^3-\frac{\omega ^2 \lp 3 (a |\nh| - i \ell) + 4 i\rp}{a^2+1} \nn\\
&&+\frac{\omega  \left(a^2 \left(3 |\nh|^2+\ell -4\right)-6 i a |\nh| (\ell -1)-(\ell-1) (3 \ell -4)\right)}{\left(a^2+1\right)^2} \nn\\
&&-\frac{(a|\nh| - i \ell) \left(a^2 \left(|\nh|^2+\ell-2\right)-2 i a |\nh| (\ell-1)-(\ell-2) (\ell-1)\right)}{\left(a^2+1\right)^3}\,.
\eeqa

For a given solution of this equation, eqs.~\eqref{cRLaguerre} and \eqref{delR} yield the profile of $\delta m(r)$ as
\beq\label{delmLag}
\delta m =e^{- \frac{r^2}{2(1+a^2)}} r^{|\nh|} L_k^{|\nh|}\lp \frac{r^2}{2(1+a^2)}\rp\,.
\eeq
Using this in the linearized perturbation equations we can obtain $\delta p_r$ and $\delta p_\phi$, which are uniquely determined once regularity at infinity is imposed. They are finite polynomials
but their general expressions are cumbersome, so we do not give them explicitly. Nevertheless, one can readily obtain the coefficients of the polynomials in particular cases by inserting the specific polynomial \eqref{delmLag} in the linearized equations of motion and solving the resulting algebraic equations for the coefficients.

The solution to the cubic \eqref{w cubic quant} for $\omega$ can be given explicitly for generic values of $\ell$ and $\nh$, but it is rather unilluminating. Instead, we will discuss generic features of axisymmetric and non-axisymmetric modes, and then consider certain special modes.

\subsubsection*{Axisymmetric modes: $\nh=0$}

\begin{itemize}
\item $k=0$, $\ell=0$. There are no non-trivial regular modes. Besides the trivial constant mode with $\omega = 0$ we find
\begin{equation}
	 \omega = \frac{2}{1+a^2}(i \pm a)\,,
\end{equation}
which appear to be unstable modes, but their profiles for $p_\phi$ 
approach a constant at infinity, which results in unphysical infinite angular momentum.

\item
$\kh \geq 1$, $\ell=2,4,6,\dots$. All the solutions to \eqref{w cubic quant} yield regular profiles of the form
\beqa
\delta p_r &=& e^{- \frac{r^2}{2(1+a^2)}} r \sum_{i = 0}^{\kh}  \delta p_r^{(i)} r^{2 i}\,, \\
\label{pert ansatz 3}
	\delta p_\phi &=& e^{- \frac{r^2}{2(1+a^2)}} r^{2} \sum_{i = 0}^{\kh}  \delta p_\phi^{(i)} r^{2 i}\,.
\eeqa

\end{itemize}

\subsubsection*{Non-axisymmetric modes: $|\nh| \geq 1$}

\begin{itemize}

\item

$\kh = 0$,  $\ell=|\nh|$. The frequencies of these fundamental quasinormal `bar-modes' obtained from \eqref{w cubic quant} are
\beqa
\label{w0 wpm}
	\omega_0 &=& \frac{(|\nh|+2) a - i (|\nh|-2)}{1+a^2}, \\
\omega_\pm &=& \frac{\sqrt{|\nh|-1}}{1+a^2} \lp  a \sqrt{|\nh|-1} \pm 1 - 
	i \lp \sqrt{|\nh|-1} \mp a \rp \rp\,.
\eeqa
Modes with $\omega = \omega_0 $ have momenta which are regular for $|\nh|>2$, 
and they are stable. For $|\nh|=1,2$ they are singular, \ie\ unphysical.

Modes with $\omega = \omega_\pm $ are regular for $|\nh| \geq 1$ and have profiles of the form
\beqa
\delta p_r &=& e^{- \frac{r^2}{2(1+a^2)}} r^{|\nh|+1}\delta p_r^{(1)}\,, \\
	\delta p_\phi &=& e^{- \frac{r^2}{2(1+a^2)}} r^{|\nh|+2} \delta p_\phi^{(1)}\,.
\eeqa
The mode $\omega_+$ with $|\nh|=2$ and $a=1$ is the corotating black bar mode \eqref{blackbarmode}. More generally, the modes $\omega_+$ with $|\nh|\geq 2$ and $\omega_+=a=\sqrt{|\nh|-1}$ are purely real and correspond to the non-axisymmetric corotating zero-modes in \eqref{MPnaxizero}.

\item $k\geq 1$. All profiles are regular, with momenta of the form
\beqa\label{pert ansatz 2}
\delta p_r &=& e^{- \frac{r^2}{2(1+a^2)}} r^{|\nh|-1} \sum_{i = 0}^{\kh+1}  \delta p_r^{(i)} r^{2 i}\,, \\
\label{pert ansatz 3}
	\delta p_\phi &=& e^{- \frac{r^2}{2(1+a^2)}} r^{|\nh|} \sum_{i = 0}^{\kh+1}  \delta p_\phi^{(i)} r^{2 i}\,.
\eeqa
When written in Cartesian coordinates these modes are manifestly regular at the origin, 
even when $|\nh|=1$, for which the radial profiles behave as $p_r \sim 1$,  $p_\phi \sim r$ near $r=0$. 
The contribution from the angular part $e^{i \nh \phi}$ plays a crucial role for regularity. 
\end{itemize}

These modes can become unstable for sufficiently large values of $a$, as we discuss below.

\paragraph{Schwarzschild modes.} When $a=0$ the solutions to \eqref{w cubic quant} are
\begin{equation}
\label{w schw}
	\omega_{\pm}^{\textit{Sch}} = \pm \sqrt{\ell -1} - i (\ell - 1), \qquad  \omega_{0}^{\textit{Sch}} = - i (\ell - 2)\,.
\end{equation}

Modes with frequencies $\omega_{\pm}^{\textit{Sch}}$ are physical (have finite total angular momentum) for $\ell > 1$. This matches the earlier result of \cite{Emparan:2014aba} for the quasinormal frequencies that are scalars of $S^{D-2}$ for the Schwarzschild solution at large $D$.

Modes with frequency  $\omega_{0}^{\textit{Sch}} $ are regular only for $\ell > 2$. They can be seen to have constant $\delta \cR$, which identifies them as vector deformations of the $S^{D-2}$. The calculation of \cite{Emparan:2014aba} gave the vector frequency as $\omega_0=-i(\ell-1)$. The difference with \eqref{w schw} is simply due to the fact that in app.~\ref{app:sphhar} we identified $\ell$ using the scalar spherical harmonics, while for the vector harmonics $\ell$ is shifted by 1.
 
All the allowed modes in \eqref{w schw} are stable, in agreement with the proven mode stability of the Schwarzschild-Tangherlini solution in all $D$ \cite{Ishibashi:2003ap}.

\paragraph{Near-critical unstable modes.} 
For 
\begin{equation}\label{corotzmodes}
	\qquad a=a_c \equiv\sqrt{\ell-1}\,, \qquad  \omega = |\nh| \frac{a}{1+a^2} \,,\qquad \ell\geq 2\,,
\end{equation}
we recover the corotating zero modes discussed in section \ref{sec:corotating}. These modes have purely real frequency, but they mark the appearance of unstable modes as $a$ increases past each critical value $a_c$. We can verify this by moving slightly away from the critical points, by setting
\beq
\omega=|\nh| \frac{a_c}{1+a_c^2}+\delta\omega\,,\qquad a=a_c+\delta a\,.
\eeq
Linearizing \eqref{w cubic quant} in $\delta\omega$ and $\delta a$ we find that the frequency develops an imaginary part,
\beq\label{ImOm}
\textrm{Im}\,\omega =\delta a\,\frac{a_c}{a_c^2(1+a_c^2)^2+\nh^2}\lp \frac{2\nh^2 }{1+a_c^2}+a_c^4-1\rp \,.
\eeq
Since $a_c\geq 1$, we see that $\textrm{Im}\,\omega>0$ whenever $\delta a>0$, so the mode is unstable, while if $\delta a<0$ the mode has $\textrm{Im}\,\omega<0$ and therefore is stable. Hence, as the rotation increases crossing each of the critical values, a new unstable mode is added to the MP black hole.

Eq.~\eqref{ImOm} gives the growth rate of the bar-mode instability near the threshold ($a_c=1$ and $\nh=2$) as
\beq\label{ImOmbar}
\textrm{Im}\,\omega=\frac12 \delta a\,.
\eeq
Interestingly, this unstable growth rate of bar modes has been computed numerically in $D=6,7$ in \cite{Shibata:2010wz} and \cite{Dias:2014eua}, who find (in units where $r_0=1$)
\beq
\textrm{Im}\,\omega  \sim C_\tau \delta a
\eeq
with
\beqa
&&C_\tau\sim 0.51\quad (D=6)\qquad  C_\tau\sim 0.54\quad (D=7)\qquad \cite{Shibata:2010wz}\,,\\
&&C_\tau\sim 0.521\quad (D=6,7)\qquad \cite{Dias:2014eua}\,.
\eeqa
The leading-order large-$D$ result from \eqref{ImOmbar},
\beq
C_\tau=1/2
\eeq
is in agreement with the numerical calculations to a few percent level. More generally one can readily verify that the plots of quasinormal frequencies, both real and imaginary, obtained from \eqref{w cubic quant} as a function of $a$ agree very well with the results presented in \cite{Dias:2014eua}.

\paragraph{Bar modes and the CFS instability.} 

The real part of the frequency of the bar mode near the critical point is
\beq\label{reOm}
\textrm{Re}\,\omega =|\nh|\Omega-\frac14 \delta a\,|\nh|\,,
\eeq
so we see that as the rotation increases, the stable mode before the critical rotation (with $\delta a<0$), rotates faster than the black hole, whereas the unstable mode (with $\delta a>0$) rotates more slowly than the black hole. The superradiant limit $\omega =|\nh|\Omega$ corresponds of course to exact corotation.

This behavior is strongly reminiscent of the Chandrasekhar-Friedman-Schutz (CFS) instability in neutron stars \cite{Chandrasekhar:1992pr,Friedmann:1978}: unstable modes are present only for perturbations that rotate in the same sense that the star but more slowly than it. This is because when the perturbation moves backwards relative to the star, but forwards relative to inertial observers, it excites the emission of gravitational waves that remove positive angular momentum from the mode, driving the deformation even slower. In our set up the emission of gravitational waves is suppressed, so the details of the instability mechanism are not the same, but it seems plausible that the two phenomena are related.

\section{Corotating perturbations of black bars}\label{sec:pertbars}

Black bars approach black strings as $\Omega\to 0$, so it is natural to expect that at sufficiently small $\Omega$ they develop instabilities similar to the Gregory-Laflamme instability of black strings. This argument, however, does not determine at what values of $\Omega$ the instabilities set in. This requires a perturbative analysis of finite black bars with non-zero $\Omega$.

The generic linearized perturbations of black bars with $0<\Omega<1/2$ are rather more complicated than those of either the black strings or the MP black holes. Nevertheless, using \eqref{stateq2} we have been able to explicitly obtain corotating, zero mode perturbations. These appear at discrete values of $\Omega$, and approach the GL zero modes of a black string as $\Omega\to 0$. As with the MP black holes, we expect that these zero modes mark the addition of new unstable modes, as well as indicate new branches of stationary, `bumpy black bar' solutions.

Using the Cartesian coordinates of \eqref{cartcorot}, the equation for corotating linear perturbations $\dcR$ is
\beq
\nabla^2 \dcR- \lp \frac{x}{\ell_\perp^2} \pd_x  +\frac{y}{\ell_\|^2} \pd_y  \rp \dcR+\dcR =0\,.
\label{eq:pertEqCart}
\eeq
We look for factorized solutions
\beq
\dcR=f_x(x)f_y(y)\,.
\eeq
These must satisfy
\beqa
\lp \pd_x^2 -\frac{x}{\ell_\perp^2} \, \pd_x +1 \rp f_x &= \frac{\ny}{\ell_\|^2} f_x\,,\\
\lp \pd_y^2-\frac{y}{\ell_\|^2}\pd_y  \rp f_y&=-\frac{\ny}{\ell_\|^2}  f_y\,,
\eeqa
where, for later convenience, we have written the separation constant as $\frac{\ny}{\ell_\|^2}$.

These equations are again of confluent hypergeometric type. If we demand regular, algebraically-bounded behavior at the irregular point at infinity, the solutions are Hermite polynomials
\beq
f_x(x)=  H_\nx \lp\frac{x}{\sqrt{2}\ell_\perp}\rp\,, \qquad
 f_y(y)=  H_\ny \lp\frac{y}{\sqrt{2}\ell_\|}\rp\,,
\eeq
where $\nx$ is another constant given in terms of $n_y$ and $\Omega$ by
\beq
\nx= \ell_\perp^2-\frac{\ell_\perp^2}{\ell_\|^2}\ny\,.
\label{eq:defny}
\eeq

The solutions are finite polynomials only if $\nx$ and $\ny$ take non-negative integer values. Therefore, co-rotating zero modes exist for a discrete set of values of $\Omega$ determined by solving \eqref{eq:defny}. This gives
\beq\label{Ommodes}
\Omega=\frac{\sqrt{1-\nx} \sqrt{\ny-1}}{\left|\ny-\nx\right|}
\eeq
 (recall we only consider $\Omega\geq 0$).

When $\Omega=0$, which is a static, infinite black string and is obtained for either $\nx=1$ or $\ny=1$, we do not obtain anything new. The perturbation with $\nx=1$ is simply a translation of the black string in the orthogonal direction $x$, while $\ny=1$ are sinusoidal deformations along $y$ corresponding to the GL zero modes of a black string.

There is only one mode that is not constant along the orthogonal direction $x$, \ie\ with $\nx\neq 0$, namely, $\nx=2$ and $\ny=0$. However, this is the bar-mode along $x$ of the $\Omega=1/2$ MP black hole (\ie\ \eqref{blackbarmode}), so again we do not get any new physical solution.

The remaining modes all have $\nx=0$ and thus are uniform along $x$. They have
\beq
\Omega=\frac{\sqrt{\ny-1}}{\ny}\,,\qquad \ell_\|^2 = \ny\,,\qquad \ell_\perp^2=\frac{\ny}{\ny-1}\,,
\eeq
and
\beq
\dcR =   H_{\ny} \lp \frac{y}{\sqrt{2\ny}}\rp \,.
\eeq
For $\ny=2$ this is again a bar-mode of the $\Omega=1/2$ MP black hole, so we disregard it. However, for
\beq\label{goodmodes}
\ny=3,4,5,\dots\,,\qquad \Omega=\frac{\sqrt{2}}{3}\,,\frac{\sqrt{3}}{4}\,,\frac{2}{5}\,\dots
\eeq
we find genuinely new zero modes, which extend all the way down to $\Omega=0$ as $\ny\to\infty$. Remarkably, these are the same values \eqref{MPnaxizero} of $\Omega$ for which the MP black holes admit zero modes (the one at $\Omega=1/2$ gives the black bar itself), even if the spins of the corresponding solutions are different (see fig.~\ref{fig:OmJ}).
 
\begin{figure}[t]
	\centerline{\includegraphics[width=1\textwidth]{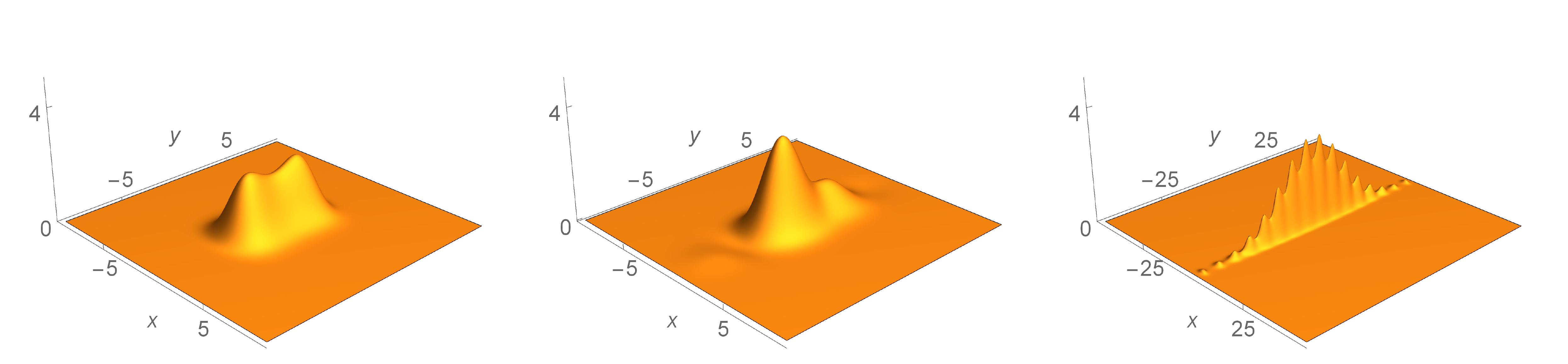}}
	\caption{\small Mass density profile $m(x,y)=\exp(\cR+\epsilon\, \dcR)$ for black bars perturbed by zero modes. From left to right: perturbations with $\ny=4,5,200$. \label{fig:barpert}}
\end{figure}

These perturbations create bumps along the length of the black bars, as shown in figure \ref{fig:barpert}.
The first non-trivial mode, with
\beq
\ny=3\,,\qquad \Omega=\frac{\sqrt{2}}3\approx 0.47\,,
\eeq
has the profile
\beq
\delta\cR=-\lp\frac{2}{3}\rp^{3/2} y\lp 9-y^2\rp\,,
\eeq
which is odd in $y$. We are not representing it in fig.~\ref{fig:barpert} since its skewed gaussian shape is not very illustrative. 
The next mode, with
\beq
\ny=4\,,\qquad \Omega=\frac{\sqrt{3}}4\approx 0.43\,,
\eeq
has the profile
\beq
\delta\cR=12-6y^2+\frac{y^4}4\,,
\eeq
and is shown in fig.~\ref{fig:barpert}~(left). 

We expect that, when continued into the non-linear regime, both the even and the odd modes give rise to branches of bumpy black bars. However, the even modes can be added or subtracted from the initial black bar, creating a dip or a rise at its center. These two possibilities will lead to two different branches of bumpy black bars, analogous to what happens for bumpy black holes \cite{Dias:2014cia,Emparan:2014pra}. In contrast, the odd modes will give only one branch of skewed bumpy black bars.

Modes with high $\ny$, \ie\ $\Omega\ll 1$, give a perturbation of the mass density that is asymptotically of the form
\beq
\delta m(x,y)=e^{\cR_{\mathrm{bar}}(x,y)}\,H_{\ny} \lp \frac{y}{\sqrt{2\ny}}\rp  \to e^{1-\frac{x^2}{2}-\frac{y^2}{4 \ny}}\cos\lp y-\frac{\ny \pi}{2} \rp\,.
\eeq
Therefore, very long black bars have zero modes that, away from the edges of the bar, where $|y|\ll \ell_\|=\sqrt{\ny}$, approach the sinusoidal oscillations of the GL modes of a black string, as we anticipated. We see that the modes with even $\ny$ converge to $\cos y$, whereas the modes with odd $\ny$ converge to $\sin y$.  In this limit, odd and even modes are equivalent as they are simply shifted in $y$ by $\pi/2$. However, for non-zero values of $\Omega$ the even and odd modes are physically distinct from each other.

The linear stability of black bars requires the investigation of non-corotating perturbations, which we have not done, but we can expect that unstable modes are added after each new zero mode appears. Given the similarities to the GL phenomenon, we find it natural to conjecture that black bars in the limit $D\to\infty$ are linearly stable when they are short enough, more precisely when
\beq
\frac{\sqrt{2}}3<\Omega<\frac12\,,
\eeq
and then only in this rather narrow range. Then, as $\Omega$ decreases below each of the critical rotation values \eqref{goodmodes}, the black bar will successively acquire new unstable modes.

\section{Final discussion and outlook}\label{sec:concl}

Effective theories of large-$D$ black holes have by now been developed in several forms, sometimes with non-coincident but overlapping ranges of application. 
The approach we have proposed in this article, which views black holes as localized bulges on a very thin black brane, in principle only deals with perturbation amplitudes $\sim 1/D$ and a range of  horizon lengths $\sim 1/\sqrt{D}$. But within this extent, we have been able to recover the most interesting physics of the Schwarzschild and Myers-Perry black holes. 
The set of equations \eqref{dyna}, \eqref{dynb}, initially designed to encode the dynamics of black branes, also manage to account for fine details of the fluctuations of an spheroidal, rotating black hole.

Black bars are another outcome of our study, and they deserve further investigation. Obtaining their unstable decay rate to GL-like inhomogeneities should be feasible, possibly analytically and if not, then numerically. But understanding their behavior at finite values of $D$ is probably more relevant. Long-lived black bars have been observed numerically in $D=6$ and $D=7$ \cite{Shibata:2010wz}, and may perhaps appear (though probably very short-lived) even in $D=5$, since black strings exist in this dimension. Pursuing the investigation of their properties through a combination of numerical and analytical methods is a rather compelling problem in higher-dimensional General Relativity.

\bigskip
\noindent \textit{The whole membrane}
\\
\noindent How does our approach relate to the large-$D$ construction in \cite{Suzuki:2015iha} of stationary black holes? The spheroidal rotating membranes of \cite{Suzuki:2015iha} describe all of the horizon of the MP black hole, with polar angles varying over the entire range $0\leq \theta\leq \pi/2$, and not just a small cap of extent $\Delta\theta=\ord{1/\sqrt{D}}$. However, as we argue in appendix~\ref{app:sphhar}, the spheroidal membrane equations are too coarse-grained to resolve the structure of the waveforms of quasinormal modes, which occurs on the shorter scales that our approach captures in detail. 

On a more technical level, the character of the differential equations is different in each of the two approaches: they are of first order in $\theta$ in \cite{Suzuki:2015iha}, and of second order in $r$ in \eqref{eqmpolar}-\eqref{eqpthpolar}. This has consequences when trying to solve them beyond linear perturbation theory. Nevertheless, the two approaches overlap on a common angular region $1/\sqrt{D}\ll\theta\ll 1$, equivalent to $1\ll r\ll \sqrt{D}$, where they can be matched to each other. It seems likely that such matching constructions, which can be readily made for the solutions with gaussian profiles, will be needed in order to better understand the complete space of solutions.

The two approaches are then seen to be complementary. In this regard, it will be interesting to obtain black bars as effective membranes extending over the entire range of polar angles. A matching construction using the solutions in this article is straightforward, but perhaps black bars can also be found entirely within the approach of \cite{Suzuki:2015iha}\footnote{In this case, the bar amplitude may have to be $\ord{1/D}$ in order to be embeddable in the background as a distorted spheroidal membrane. This problem requires membrane equations with time dependence, which are more general than those presented in \cite{Suzuki:2015iha}, but which can be obtained from the formalism in \cite{Tanabe:2015hda,Tanabe:2016pjr}, and also from \cite{Bhattacharyya:2015fdk}.}.

\bigskip
\noindent \textit{Bumpy black holes and black bars, and other directions}
\\
\noindent We have found linearized zero-mode perturbations of both MP black holes and black bars, which indicate the appearance of new branches of solutions. Bumpy black holes are indeed known at finite $D$ \cite{Dias:2014cia,Emparan:2014pra}, and other solutions such as black rings have been constructed and studied as large-$D$ membranes in \cite{Tanabe:2015hda,Tanabe:2016pjr,Mandlik:2018wnw}. Can we find these as non-linear solutions of our master equation for stationary configurations \eqref{stateq2}? Our preliminary attempts at this are inconclusive yet (see app.~\ref{app:nonlin}). Possibly, a matching to the description in \cite{Suzuki:2015iha} is needed.

Finally, our approach to large-$D$ black holes can be readily extended to investigate charged rotating black holes, as well to describe black hole collisions. Work is in progress on these problems, and we hope to report on it in the future.

\section*{Acknowledgements}
We are indebted to Kentaro Tanabe and Ryotaku Suzuki for long-standing collaboration on large $D$ black holes, with many useful discussions. This work is supported by ERC Advanced Grant GravBHs-692951 and MEC grant FPA2016-76005-C2-2-P.


\appendix

\section{Large-$D$ limit of Schwarzschild and Myers-Perry black holes}\label{app:SchwMP}

The Schwarzschild-Tangherlini and Myers-Perry black holes admit large-$D$ limits in which they fit in the class of solutions of the form \eqref{largeDmetric}, \eqref{ACG} with arbitrary $p$. We will illustrate this by casting them as gaussian bulges on strings, then on 2-branes, and finally on $2p$-branes.

\subsection{Gaussian string}

Begin with the Schwarzschild-Tangherlini solution,
\beq\label{schwn}
ds^2=-\lp 1-\frac1{\hr^n}\rp d\htt^2+\frac{d\hr^2}{1-\hr^{-n}}+\hr^2d\theta^2+\hr^2\cos^2\theta d\Omega_n\,,
\eeq
where we have set the horizon radius to one. In analogy to our discussion in the introduction, the sphere $S^{n+1}$ is built here as a fibration of spheres $S^n$ over the interval $\theta\in [-\pi/2,\pi/2]$, with the equator at $\theta=0$.

In order to take the large-$D$ limit, change $(\hr,\theta)\to(\br,z)$ as
\beq
\br=\hr\cos\theta\,,\qquad z=\sqrt{n}\,\hr\sin\theta\,,
\eeq
and let $n\to\infty$ while keeping $z$ finite. In this limit we are focusing on a small region around
\beq
\theta =\ord{ \frac1{\sqrt{n}}}\,.
\eeq

Introduce
\beq
\sR=\br^n\,.
\eeq
Then, since 
\beq
\hr^2=\br^2+\frac{z^2}{n}\,,
\eeq
we have, at large $n$,
\beq
\hr^n = \sR\, e^{z^2/2}\,.
\eeq

We are viewing  $\br$ (and $\sR$) as the coordinate orthogonal to the brane, and $z$ as the coordinate along the brane. Since the horizon is actually at constant $\hr$, this means that we regard it as a brane that is bent along the $z$ direction. 

In these coordinates, and in the limit of large $n$, the solution \eqref{schwn} becomes
\beq
ds^2\simeq -A d\htt^2+\frac1{n^2}\frac{d\sR^2}{A \sR^2}+\frac1n\lp 1+\frac1n \frac{z^2 e^{-z^2/2}}{A\sR}\rp dz^2+\br^2 d\Omega_n\,, 
\eeq
with
\beq
A=1-\frac{e^{-z^2/2}}{\sR}\,.
\eeq
Change to the Eddington-Finkelstein time $t$,
\beq
t =\htt-\frac1n\ln(A \sR)=\htt-\frac1n \ln\lp \sR -e^{-z^2/2}\rp\,.
\eeq
The solution now takes the form of \eqref{largeDmetric} with
\beq
m(z)=e^{-z^2/2}\,,\qquad p(z)=z\, e^{-z^2/2}\,.
\eeq
This limit had also been obtained (though in a different gauge) in \cite{Suzuki:2015axa}.

\subsection{Gaussian singly-rotating membrane}

If we extend now the previous limit to involve two directions on the $S^{D-2}$, we can incorporate the effects of rotation on the plane of these directions.

Consider then the Myers-Perry black hole with a single spin in $D=n+5$ \cite{Myers:1986un},
\beqa
ds^2&=&-\lp 1-\frac{1}{\hr^n \Sigma}\rp d\htt^2 +\frac{2 a \sin^2\theta}{\hr^n\Sigma}d\htt d\hph+\lp \hr^2+a^2+\frac{a^2\sin^2\theta}{\hr^n\Sigma}\rp\sin^2\theta\, d\hph^2
\nn\\
&&+\frac{\Sigma}{\Delta}d\hr^2+\Sigma d\theta^2+\hr^2\cos^2\theta d\Omega_{n+1}
\eeqa
where 
\beq
\Sigma=\hr^2+a^2\cos^2\theta\,,\qquad \Delta =\hr^2+a^2-\frac1{\hr^n}\,,
\eeq
and $\theta\in [0,\pi/2]$.

The horizon is at $\hr=r_H$, where $\Delta(r_H)=0$. Then it satisfies
\beq
r_H=\lp 1+\frac{a^2}{r_H^2}\rp^{-\frac{1}{n+2}}\,.
\eeq
When $n\to\infty$ we have
\beq
r_H\to 1\,,\qquad r_H^n\to \frac1{1+a^2}\,.
\eeq

Now change $(\hr,\theta)\to(\br,r)$, with
\beq\label{rhor}
\br =\hr\cos\theta\,,\qquad r=\sqrt{n(\hr^2+a^2)}\sin\theta\,.
\eeq
It is useful to note that
\beq
\Sigma\lp\frac{d\hr^2}{\hr^2+a^2}+d\theta^2\rp+\lp \hr^2+a^2\rp\sin^2\theta\, d\hph^2+\hr^2\cos^2\theta d\Omega_{n+1}=d\br^2+\frac{dr^2+r^2d\hph^2}{n}+\rho^2d\Omega_{n+1}\,.
\eeq
$\br$ is the coordinate orthogonal to the membrane. The membrane worldvolume is described in polar coordinates in which $r$ is the radius and $\hph$ the polar angle.

We introduce
\beq
\sR=(1+a^2)\br^n\,.
\eeq
Then
\beq
\hr^n= \frac{\sR}{1+a^2}\,e^{\frac{r^2}{2(1+a^2)}}\,.
\eeq
In the new coordinates, expanding in $1/n$, the metric becomes
\beqa
ds^2&\simeq& -Ad\htt^2+\frac1n\frac{2a}{1+a^2}\frac{r^2e^{-\frac{r^2}{2(1+a^2)}}}{\sR}d\htt\,d\hph+\frac{r^2}{n}\lp 1+\frac{r^2}{n}\frac{a^2}{(1+a^2)^2}\frac{e^{-\frac{r^2}{2(1+a^2)}}}{\sR}\rp d\hph^2\nn\\
&&+\frac{1}{n^2}\frac{d\sR^2}{A\sR^2}+\frac1n\lp 1+\frac{r^2}{n(1+a^2)^2}\frac{e^{-\frac{r^2}{2(1+a^2)}}}{A\sR}\rp dr^2+\br^2d\Omega_{n+1}\,,
\eeqa
with
\beq
A=1-\frac{e^{-\frac{r^2}{2(1+a^2)}}}{\sR}\,.
\eeq

In order to go to Eddington-Finkelstein coordinates, change
\beq
\htt=t+\frac1n\ln(A\sR)\,,\qquad \hph=\phi-\frac1{n}\frac{a}{1+a^2}\ln(A\sR)\,.
\eeq
The metric is now of the form of \eqref{largeDmetric} with
\beqa
m(r)&=&e^{-\frac{r^2}{2(1+a^2)}}\,,\nn\\
p_r(r)&=&-\frac{r}{1+a^2}e^{-\frac{r^2}{2(1+a^2)}}\,,\\
p_\phi(r)&=&\frac{a r^2}{1+a^2}e^{-\frac{r^2}{2(1+a^2)}}\nn\,.
\eeqa
This is the same as we found in \eqref{mMP}, \eqref{prMP}, \eqref{pphMP} (here with $\cR_0=0$) by direct solution of the large-$D$ effective membrane equations. Therefore the MP black hole is represented as a gaussian blob on the membrane, whose width expands as $a$ grows.

\subsection{Gaussian multiply-rotating brane}

Finally, we show how this construction extends to a more general limit of the Myers-Perry black hole, in which a gaussian bulge extends along any finite number of directions. For definiteness, we will consider that these are $2p$ directions, with $p$ spins turned on along corresponding rotation planes. Note, however, that it is not necessary that the rotation is non-zero in order to find a gaussian bulge on the brane: if there is no rotation, the bulge is just narrower. Also, we could easily add a single, odd direction to the profile, \ie\ the brane need not be an even-brane but for simplicity we will not do it.

Setting $D=2p+n+3$, the metric is \cite{Myers:1986un}
\begin{align}
ds^2=&-dt^2+\frac{ \hat{r}^\gamma}{\Pi F}\left(dt+\sum_{i=1}^{p}a_i\hat{\mu}_i^2\,d\phi_i\right)^2+\frac{\Pi F}{\Pi-  \hat{r}^\gamma}d\hat{r}^2\nn\\
+&\sum_{i=1}^{p}(\hat{r}^2+a_i^2)(d\hat{\mu}_i^2+\hat{\mu}_i^2d\phi_i^2)+d\rho^2+\rho^2 d\Omega^{2}_{n+1}\,,
\label{eq:separatedMetric}
\end{align}
where $\gamma=2\,(1)$ if $D$ is odd (even), the metric functions are
\begin{align}
F&=1-\sum_{i=1}^{p}\frac{a_i^2\mu_i^2}{\hat{r}^2+a_i^2}\,,\\
\Pi&=\hat{r}^{n}\prod_{i=1}^{p}\left(\hat{r}^2+a_i^2\right)\,,
\end{align}
and the coordinates satisfy implicitly
\begin{align}
\rho^2=\hat{r}^2\left(1-\sum_{i=1}^{p}\hat{\mu}_i^2\right)\,.
 \label{eq:decompSpherRadial}
\end{align}
The horizon is at $\hat{r}=r_H$ defined by $\left.\Pi\right|_{r_H}=r_H^\gamma$, which implies that
\begin{align}
r_H=\left(\prod_{i=1}^{p}\left(1+\frac{a_i^2}{r_H^2}\right)\right)^{-\frac{1}{n+2p-\gamma}}\xrightarrow{n\gg 1}1\,,
\end{align}
and
\beq
r_H^n\to \mathrm{P}\equiv \prod_{i=1}^{p} \left( 1+ a_i ^2 \right)\,.
\eeq

Now we focus on regions where $\hat{\mu}_i\simeq O\left(n^{-\frac{1}{2}}\right)$, so we define the new variables 
\begin{align}
r_i = \sqrt{n\left( \hat{r}^2+a_i^2 \right)}\hat{\mu}_i\,.
 \end{align}
Enforcing the constraint \eqref{eq:decompSpherRadial}, we note that, to leading order,
\begin{align}
&\frac{\Pi F}{\Pi-  \hat{r}^\gamma}d\hat{r}^2+d\rho^2+
\sum_{i=1}^{p}(\hat{r}^2+a_i^2)(d\hat{\mu}_i^2+\hat{\mu}_i^2d\phi_i^2)\nn\\
\simeq\,&\,d\rho^2+\sum_{i=1}^{p}\frac{dr_i^2+r_i^2d\phi_i^2}{n}\,.
\end{align}
We introduce
\begin{align}
\sR=\rho^n \mathrm{P}\,.
\end{align}
Then
\begin{align}
\hat{r}^n
&=\frac{\sR}{\mathrm{P}}\left(1-\frac{1}{n}\sum_{i=1}^{p}\mu_i^2\right)^{-\frac{n}{2}}\simeq \frac{\sR}{\mathrm{P}}\left(1+\frac{1}{n}\sum_{i=1}^{p}\mu_i^2\right)^{\frac{n}{2}}\nn\\
&=\frac{\sR}{\mathrm{P}}\left(1+\frac{1}{n}\sum_{i=1}^{p}\frac{r_i^2}{\hat{r}^2+a_i^2}\right)^{\frac{n}{2}}\rightarrow \frac{\sR}{\mathrm{P}}e^{\sum_{i=1}^{p}\frac{r_i^2}{2\left(1+a_i^2\right)}}\,.
\end{align}
In the new coordinates, expanding in $1/n$, the metric becomes
\beqa
ds^2&\simeq& -Adt^2
+\frac1n\sum_{i=1}^{p}\frac{a_ir_i^2}{1+a_i^2} \frac{e^{-\sum_{j=1}^{p}\frac{r_j^2}{2(1+a_j)}}}{\sR} dt\,d\phi_i+\frac{1}{n^2}\frac{d\sR^2}{A\sR^2}\nn\\
&&+\sum_{i,j=1}^{p}\frac{r_i r_j}{n}\lp\delta_{ij}+\frac{1}{n}\frac{a_ia_jr_ir_j}{(1+a_i^2)(1+a_j^2))}\frac{e^{-\sum_{k=1}^{p}\frac{r_k^2}{2(1+a_k^2)}}}{R} \rp d\phi_id\phi_j\nn\\
&&+\sum_{i,j=1}^{p}\frac1n\lp \delta_{ij}+\frac{r_i r_j}{n(1+a_i^2)(1+a_j^2)}\frac{e^{-\sum_{k=1}^{p}\frac{r_k^2}{2(1+a_k^2)}}}{A\sR}\rp dr_i dr_j\nn\\
&&+\br^2d\Omega_{n+1}\,,
\eeqa
with
\beq
A=1-\frac{e^{-\sum_{i=1}^{p}\frac{r_i^2}{2(1+a_i^2)}}}{\sR}\,.
\eeq
With a final change to Eddington-Finkelstein coordinates,
\beq
\htt=t+\frac1n\ln(A\sR)\,,\qquad \hph_i=\phi_i-\frac1{n}\frac{a_i}{1+a_i^2}\ln(A\sR)\,.
\eeq
the metric now takes the form of \eqref{largeDmetric} with the gaussian profiles
\beqa
m(r_i)&=&e^{-\sum_{i=1}^{p}\frac{r_i^2}{2(1+a_i^2)}}\,,\nn\\
p_{r_i}(r_i)&=&-\frac{r_i}{1+a_i^2}e^{-\sum_{i=1}^{p}\frac{r_i^2}{2(1+a_i^2)}}\,,\\
p_{\phi_i}(r_i)&=&\frac{a r_i^2}{1+a_i^2}e^{-\sum_{i=1}^{p}\frac{r_i^2}{2(1+a_i^2)}}\nn\,.
\eeqa

\section{Spheroidal harmonics at large $D$}\label{app:sphhar}

Here we analyze scalar spheroidal harmonics at large $D$ and relate them to our study of quasinormal perturbations of gaussian black-hole lumps in sec.~\ref{sec:qnms}. Initially we follow appendix~C of \cite{Suzuki:2015iha}, but then we depart from it so as to highlight the differences and connections between their approach and ours.\footnote{Related aspects have been analyzed by K.~Tanabe, to whom we are indebted for private communications.}

As in \cite{Suzuki:2015iha}, we study the massless scalar field equation
\beq
\Box \Psi=0
\eeq
in flat space in $D=n+5$ dimensions, written in spheroidal coordinates,
\beqa
ds^2&=&-dt^2+(\hr^2+a^2\cos^2\theta)\lp\frac{d\hr^2}{\hr^2+a^2}+d\theta^2\rp\nn\\
&&
+(\hr^2+a^2)\sin^2\theta\, d\phi^2 +\hr^2\cos^2\theta\, d\Omega_{n+1}
\eeqa
(appropriate for embedding a rotating, MP-type, large-$D$ membrane). 

We separate variables as
\beq
\Psi=e^{-i\omega t}e^{i \nh\phi}\psi(\hr) S(\theta)\,,
\eeq
where, in order to avoid inessential details, we are assuming no dependence on the angles of the sphere $S^{n+1}$. Introducing a separation constant $\Lambda$, we obtain the equations\footnote{The differences with \cite{Suzuki:2015iha} are a shift in the definition of $\Lambda$ and a corrected typo in their radial equation, both inconsequential to the rest of the analysis.}
\beq
\lp \frac1{\hr^{n+1}}\frac{d}{d\hr} (\hr^2+a^2)\hr^{n+1}\frac{d}{d\hr}  +\frac{\nh^2 a^2}{\hr^2+a^2}+\omega^2\hr^2-\Lambda\rp\psi(\hr)=0
\eeq
and
\beq\label{Seqn}
\lp \frac{d^2}{d\theta^2}+\lp\cot\theta-(n+1)\tan\theta\rp\frac{d}{d\theta}-\frac{\nh^2}{\sin^2\theta}+\omega^2 a^2\cos^2\theta +\Lambda\rp S(\theta)=0\,.
\eeq
When $a=0$ we get the usual scalar spherical harmonics, with the separation constant quantized as
\beq
\Lambda =\ell (\ell+n+2)\,,
\eeq
where $\ell$ is a non-negative integer. This result indicates that if we consider $\ell=\ord{1}$, then we must have $\Lambda =\ord{n}$. The authors of \cite{Suzuki:2015iha} argue that it is appropriate to extend this behavior to $a\neq 0$ and set
\beq
\Lambda =n\ell +\ord{1}\,.
\eeq
They then proceed to take the the limit $n\to\infty$ in \eqref{Seqn} while keeping $\theta=\ord{1}$. This yields, to leading order,
\beq\label{Seqn2}
\frac{dS(\theta)}{d\theta}=\frac{\ell}{\tan\theta} S(\theta)\,,
\eeq
which is solved by 
\beq\label{Ssin}
S_\ell(\theta)=\sin^\ell\theta\,.
\eeq

Crucially, note that in the limit from \eqref{Seqn} to \eqref{Seqn2} the latter has become a first order equation ---so the condition of regularity at $\theta=\pi/2$, which leads to the quantization of $\ell$, is mysteriously absent---, and its solutions $S_\ell(\theta)$ have lost all the characteristic structure of the spherical harmonics with nodes in the angular direction. These two features are intimately related, and point to the fact that when $n$ is large, the angular structure of the spherical harmonics is hidden within a small region $\theta=\ord{1/\sqrt{n}}$, which is invisible when we consider $\theta=\ord{1}$.

In order to reveal this fine structure, we first rescale the angle in the by now familiar manner (cf.~\eqref{rhor}),
\beq
\theta=\frac{r}{\sqrt{n(1+a^2)}}\,,
\eeq
so that, when we now take $n\to\infty$, \eqref{Seqn} becomes
\beq
\lp \frac{d^2}{dr^2} + \left( \frac{1}{r} - \frac{r}{1+a^2} \right)\frac{d}{dr} +  \frac{\ell}{1+a^2}  - \frac{\nh^2}{r^2}\rp S(r)=0\,.
\eeq 
This is a second-order equation, exactly the kind of confluent hypergeometric equation that we encounter when we study perturbations of the gaussian MP black holes in sec.~\ref{sec:qnms}. Requiring regularity at both $r=0$ and $r\to\infty$, its solutions are given in terms of associated Laguerre polynomials,
\beq\label{SLaguerre}
S(r)= r^{|\nh|} L_k^{|\nh|}\lp \frac{r^2}{2(1+a^2)}\rp\,, 
\eeq
where the non-negative integer index $k$ specifies the quantization condition on $\ell$ through the relation \eqref{ellkm}. 

Now the eigenfunctions \eqref{SLaguerre} have the expected $k$ nodes away from $r=0$. Moreover, since the $L_k^{|\nh|}$ are polynomials of $k$-th order, then at large values of $r$ we have
\beq
S(r)\sim r^\ell\,,
\eeq
which correctly matches the behavior of \eqref{Ssin} at small $\theta$.

We conclude that our approach to localized black holes based on the effective black brane equations is able to accurately capture the detailed structure of linear perturbations of a black hole, and in particular its quasinormal modes ---not only the frequency spectrum, but also their waveforms. It can be smoothly continued into the approach of \cite{Suzuki:2015iha} at larger angles $\theta=\ord{1}$, by asymptotic matching over a common region where $1\ll r\ll \sqrt{n}$, \ie\ $1/\sqrt{n}\ll \theta\ll 1$.

\section{Large $r$ behavior and other non-linear solutions}\label{app:nonlin}

All the exact non-linear solutions for $\cR$  in secs.~\ref{sec:axisymm} and \ref{sec:bars} behave at large $r$ like 
\beq
\cR \sim -r^2\,.
\eeq
Equivalently, the mass density $m$ falls off like a gaussian at asymptotic infinity. However, the non-trivial perturbations of these solutions all tend at large $r$ to higher powers, $\delta\cR\sim r^\ell$ with $\ell>2$. We may ask whether the master equation \eqref{stateq2} admits non-linear solutions with this higher-than-quadratic behavior as $r\to\infty$.

It is not difficult to prove that this is not possible. Assume that $\cR$ extends towards $r\to\infty$ with $-\cR$ growing larger than $r^2$. Any possible dependence on $\phi$ yields subdominant effects, so we can consider \eqref{stateqpol} and drop other negligible terms to find
\beq
\cR''+\frac12 {\cR'}^2+\cR\simeq 0\,.
\eeq
This is the same as the equation in \cite{Emparan:2015hwa} governing stationary black strings. It can be integrated once to cast it as the mechanics of a one-dimensional classical particle,
\beq
\frac12 {\cR'}^2+V(\cR)\simeq 0
\eeq
in the potential
\beq
V(\cR)=\cR + C e^{-\cR}\,.
\eeq
The integration constant $C$ can be either positive or negative. If $C>0$ (so that $V$ has a minimum, which was the case of interest in \cite{Emparan:2015hwa} for black strings in a compact circle) the potential is bounded below and there is no possibility of $\cR$ diverging at large $r$. Thus we must consider $C<0$, for which $V(\cR)\to-\infty$ as $\cR\to-\infty$. In this case, however, there are no power-law solutions. Instead, we find that $\cR$ diverges logarithmically,
\beq\label{logdiv}
\cR\to 2\ln(r_0-r)
\eeq
at some (sufficiently large) finite radius $r_0$. 

This singularity is not obviously pathological: the mass density $m=e^\cR$ vanishes as $r\to r_0$, and if we set $m=0$ for $r\geq r_0$ then the configuration has finite total mass. At present, however, it is unclear to us whether this behavior can be smoothly extended along the horizon in order to obtain a complete, regular localized black hole at larger polar angles, in the same manner as the solutions with $\cR\sim -r^2$ do. 

The divergent behavior \eqref{logdiv} can be further understood by noting that \eqref{stateqpol} admits a 
logarithmic power series solution near an arbitrary point $r_0 >0 $ of the form
\begin{equation}\label{logseries}
	{\cal R}(r) = 2 \log u+ \sum_{i = 0} b_{[0],i} u^i + u^2 \log u \sum_{i = 0} b_{[1],i} u^i+
	 u^4 (\log u)^2 \sum_{i = 0} b_{[2],i} u^i + \ldots
\end{equation}
where $u = r_0 - r$. Higher powers of $\log u$ occur through the combination 
$u^{2 k} (\log u)^k$ times regular powers in $u$. The constant term $b_{[0],0}$ is arbitrary. 

Following this analysis we have obtained solutions in a large region in parameter space, which are best understood when they are smoothly connected to MP solutions that admit the corotating perturbations in sec.~\ref{sec:corotating}. They can be regarded as the non-linear extension of these zero modes. 

The perturbative construction that extends the zero modes into the non-linear regime yields corrections that diverge with $r$ more quickly the higher the perturbation order. This signals the breakdown of perturbation theory at a finite radius. 

We can also use the series \eqref{logseries} to construct numerically solutions by moving away from the singularity at $r = r_0$, integrating inwards and matching with the solutions obtained by integrating from the origin. Proceeding this way, we obtain numerical solutions which display similar behaviour to the perturbative ones, although they go large and negative even faster as $r_0$ is approached.

Given the tentative nature of these constructions, we will not give more details here. Nevertheless, we expect that, with a more detailed understanding of how to match at large $r$ these solutions to those of \cite{Suzuki:2015iha} (following app.~\ref{app:sphhar}), we should be able to construct fully non-linear bumpy black holes and black bars, and possibly other novel black hole solutions.

\end{document}